\documentclass[%
 reprint,
superscriptaddress,
 amsmath,amssymb,
 aps,
prb,
]{revtex4-2} 

\usepackage{graphicx}
\usepackage{dcolumn}
\usepackage{bm}
\usepackage{hyperref}
\usepackage[mathlines]{lineno}
\usepackage{import}

\begin{document}

\title[]{Shortest paths govern fracture nucleation in thermoset networks}

\author{Zheng Yu}
\affiliation{Department of Chemistry, University of Illinois at Urbana-Champaign, Urbana, Illinois, 61801, USA}

\author{Nick Jackson}%
 \email{jacksonn@illinois.edu}
\affiliation{Department of Chemistry, University of Illinois at Urbana-Champaign, Urbana, Illinois, 61801, USA}

\begin{abstract}
In this work, we introduce a predictive approach for determining fracture nucleation in thermosets based on shortest paths (SPs) of the network topology. This method enumerates SP sets in networks with periodic boundary conditions, with applications to both all-atom and coarse-grained simulations. We find that network fracture is most likely to nucleate on the first (shortest) SP and the strain at nucleation is linearly correlated with the topological path length. Subsequent fracture events are dictated by the instantaneous SP of partially broken networks. We further quantify the length scale dependence of SP distributions, introducing a means of bridging simulated and experimental fracture profiles.
\end{abstract}




\maketitle

Polymer networks are ubiquitous, constituting elastomers, gels, and thermosetting plastics.\cite{guPolymerNetworksPlastics2020,danielsenMolecularCharacterizationPolymer2021} The mechanism of how these networks fracture—critical for applications ranging from structural engineering to biomedical devices and stretchable electronics—continues to be a subject of intense research.\cite{rogersMaterialsMechanicsStretchable2010,zhongQuantifyingImpactMolecular2016,cretonFractureAdhesionSoft2016,kimFractureFatigueFriction2021,tauberStretchyDisorderedUnderstanding2022,aroraCoarseGrainedSimulationsFracture2022,wangFacileMechanochemicalCycloreversion2023} While crack propagation has been extensively studied,\cite{lake1967strength,livneNearTipFieldsFast2010,kinlochFractureBehaviourPolymers2013,linFracturePolymerNetworks2020,dengNonlocalIntrinsicFracture2023,maoRupturePolymersChain2017} the initiation process of fracture through bond scission within these amorphous networks is fundamental to the performance and longevity of polymeric materials, but remains poorly understood.\cite{bootsQuantifyingBondRupture2022,m.elderMechanicsNanovoidNucleation2018} Central to this fact is the need to identify the primary triggers for bond scission within amorphous networks; is fracture nucleated from localized defects, such as dangling bonds and short loops, or does the global network topology play a role?\cite{lorenzoNanostructuralHeterogeneityPolymer2015,griffith1921vi,aroraFracturePolymerNetworks2020,barneyFractureModelEndlinked2022,zhangPredictingFailureLocations2024} Clarifying this mechanism is essential to tailor material design, enabling the development of polymer networks with optimized failure resistance, durability, and degradability.\cite{zhaoSoftMaterialsDesign2021,postReviewPotentialLimitations2020}

\begin{figure}
  \centering
  \includegraphics[width=1\linewidth]{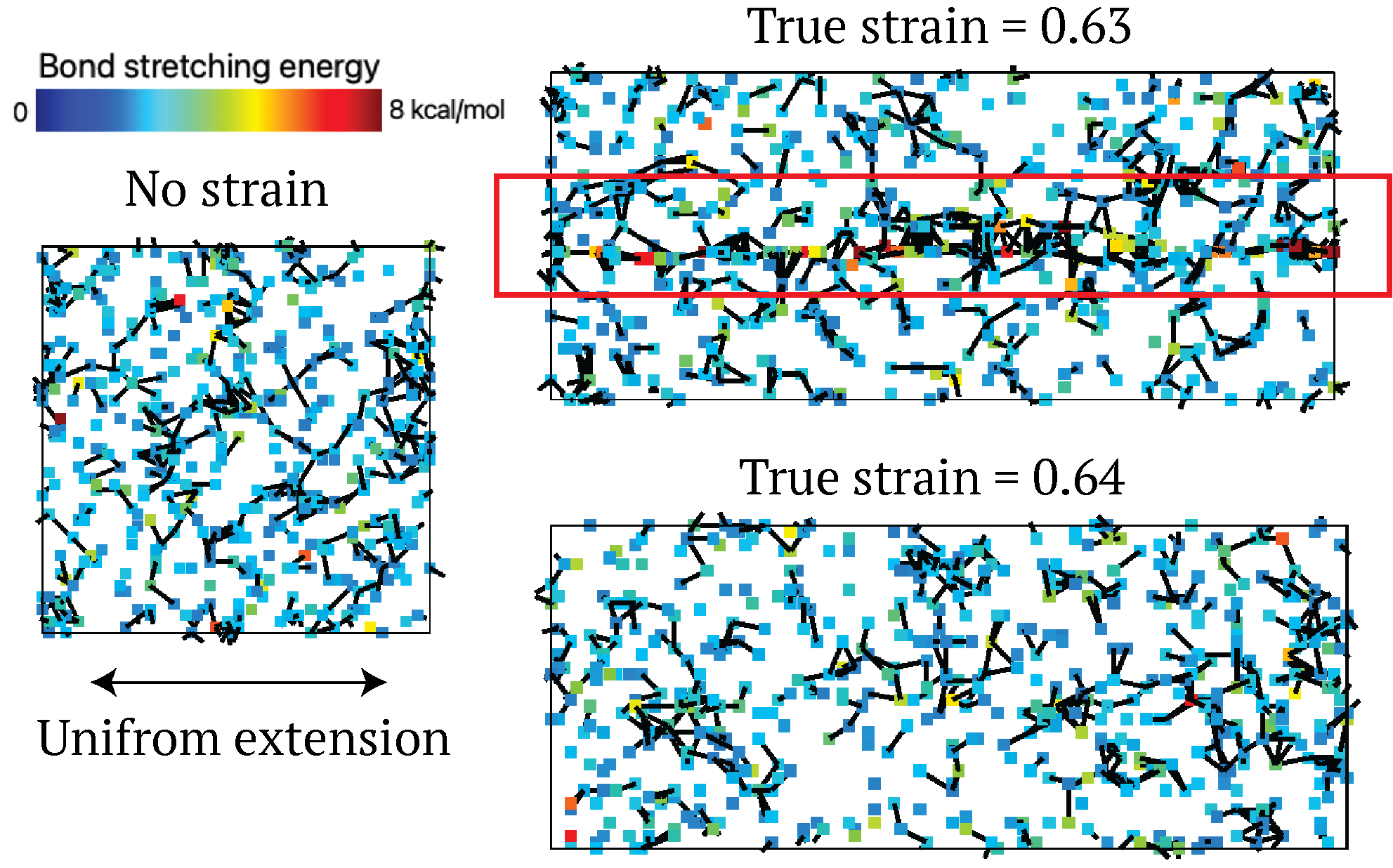}
  \caption{Local stress distributions in all-atom MLABT simulations at various strains (top view of 3D visualization). The colored points are highly stretched bonds (bond energy$>$2 kcal/mol) with color representing the bond stretching energy. Line segments are drawn if two highly stretched bonds are proximate ($<$6 {\AA}, about half the length of a monomer), to indicate stressed local regions. Initially, these regions are scattered throughout the system. As the system is deformed, the stressed regions form a nearly linear connected pathway across the simulation box along the direction of deformation before a bond breakage occurs. The pathway disappears immediately following bond scission due to stress release.}
  \label{fig:local_stress}
\end{figure}

To discern whether fracture nucleation originates from global topology or local defects, we perform thermoset fracture simulations using the Machine Learning-based Adaptable Bonding Topology (MLABT) method, which achieves quantum chemically accurate bond scission at a computational cost near that of classical molecular dynamics (MD) simulations.\cite{yuMachineLearningQuantumchemical2023,yuExploringThermosetFracture2024} Details of the computational methods are provided in Sec. S1B of the Supplementary Material.\cite{jorgensenDevelopmentTestingOPLS1996,LAMMPS,stevensInterfacialFractureHighly2001,kremerDynamicsEntangledLinear1990,zhaoUncoveringMechanismSize2022,boerner2023access} The local stress distributions indicated by the bond stretching energy visualization are investigated during uniaxial tensile deformation, as shown in Fig. \ref{fig:local_stress}. Each colored point represents a stressed bond, with black line segments indicating the proximity of two stressed bonds within a 6 {\AA} distance, thereby marking a region of concentrated stress. Prior to deformation, these regions are scattered throughout the system,  but as the network is deformed near the fracture nucleation (first bond breakage), a pattern emerges: stressed regions align to form a nearly linear connected path across the material in the direction of extension. This stress pathway vanishes immediately following bond breakage, and only reappears immediately prior to subsequent bond breakages (shown in Fig. S14). These qualitative observations are robust across MLABT MD trajectories we analyze. The emergence of these linear pathways prior to fracture suggests that the entire polymer path is taut and under maximum stress, pointing to nonlocal effects, in alignment with speculation from recent experimental measurements of fracture toughness in crack growth.\cite{dengNonlocalIntrinsicFracture2023} Therefore, our molecular simulations suggest that fracture nucleation at the molecular level is dictated primarily by global topological paths, specifically the shortest paths, and not by local defects. 

The shortest path (SP) concept, rooted in graph theory, has played a critical role in many fields ranging from matter and energy transportation to information communication.\cite{newmanNetworks2018} In the context of polymer networks, SPs are instrumental in identifying the most efficient pathways for stress transfer, offering fundamental insights into material properties through computational studies. For instance, Stevens applied SP analysis (referred to as minimal paths in his study) between two fixed walls to examine interfacial fracture in highly crosslinked coarse-grained (CG) polymer networks.\cite{stevensInterfacialFractureHighly2001} Furthermore, primitive path analysis, which builds on SPs under constraints of surrounding chains is useful to elucidate the rheology of entangled polymer liquids.\cite{everaersRheologyMicroscopicTopology2004} More recently, Cai and colleagues have demonstrated the significance of global SP lengths in understanding bond breaking and self-healing capabilities in elastomers.\cite{yinTopologicalOriginStrain2020,yinNetworkEvolutionControlling2024} Despite these advances, there is a lack of an explicit definition of SPs for networked materials in periodic simulations. Such a definition is crucial for establishing correlations with the molecular-level fracture properties, especially in thermosetting networks characterized by physically short strands (edges) and slow thermal relaxation of structures. 

In this work, we propose a novel approach to define and investigate the SP set in polymer networks, especially within theoretical frameworks employing periodic boundary conditions (PBC). The use of PBC in amorphous polymer structures ensures consistent physical constraints, even when the material undergoes deformation. Our methodology specifically targets the SPs that connect identical atoms across different periodic images, thereby resulting in a set of SPs instead of a single SP. The process for identifying the SP set comprises three steps.  
1) Generation of multi-image networks. A series of graphs containing multiple periodic images along the direction of extension are generated $\{G_p|p = 1,2,...\}$, where $p$ is the total number of images minus 1. $p>1$ is necessary because SPs can have a wavelength of more than one image, as illustrated in Fig. S1A. 2) Searching SPs using Dijkstra's algorithm between pairs of identical particles from the most distant images,\cite{dijkstraNoteTwoProblems1959}, i.e. $SP(u_i^0,u_i^p ; G_p)$, where $u_i^0$ is a particle (node) in the $0$th periodic image, and $u_i^p$ the same particle in the $p$th image in $G_p$. This step is repeated for every particle and every multi-image network, aggregating all potential SPs into the collection $\{{SP}(u_i^0,u_i^p ; G_p) | i = 1,2,...N, p = 1,2,... \}$, where $N=|u|$ is the total number of particles. 3) Elimination of redundant SPs, such as duplicate paths that differ only by starting particles and SPs at larger $p$ that repeat those found at small $p$. This step is critical for establishing a finite $p$ value, ensuring no unique SPs are overlooked in networks with large $p$. We detailed the complete methodology in the Supplementary Material Sec. S1, including the normalization of SP length, $D^g$, (topological length or geometric distance) by $p$ for paths spanning multiple images. 

Our investigation applies this SP identification strategy to three molecular models: an all-atom model simulated with the MLABT method, a traditional bead-spring CG model, and a simple network model. As the target of the work is molecular scale fracture nulceation, we focus on results from the all-atom model, while results from the other models are provided in the Supplementary Material. To streamline network analyses in the all-atom model, we treat crosslinkers as nodes, with edges representing direct connections between crosslinkers without intermediate nodes. This approach provides a framework for studying SPs in polymer networks, offering insights into their structural and mechanical properties. 

\begin{figure}
  \centering
  \includegraphics[width=0.9\linewidth]{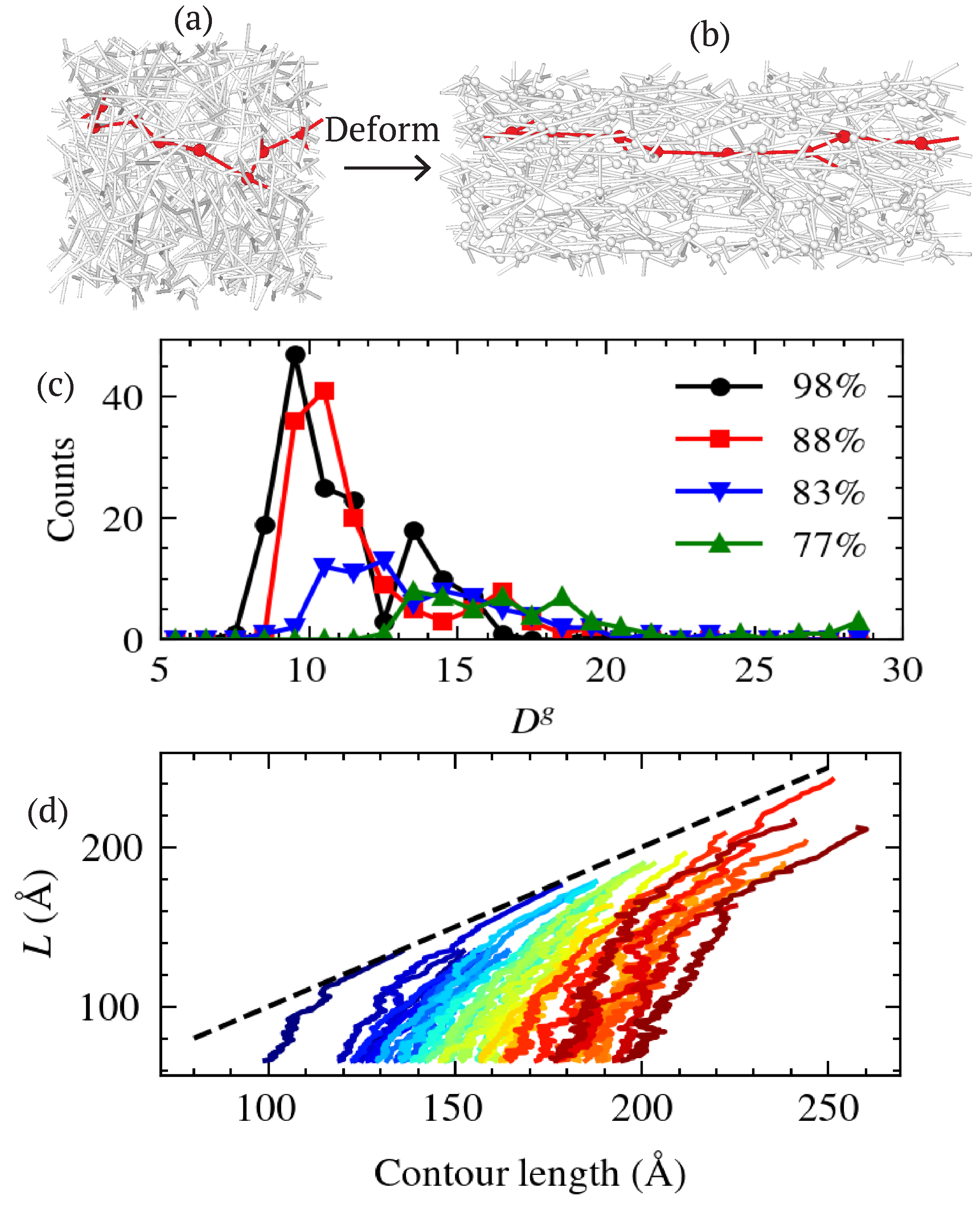}
  \caption{(a) A shortest path (SP) in the initial state from the MLABT simulation trajectory. (b) The SP is stretched taut under deformation immediately before breakage. (c) Distributions of SP lengths, $D^g$, at 3 different degrees of crosslinking in the all-atom model. (d) Contour length (the product of $D^g$ and average edge length $\bar{l}$) evolution of SPs under deformation. The dashed line represents a state of SP straightness, i.e. the contour length and the system length in the deformed direction $L$ are equal. Each line represents one SP in the set with color representing its ordering sorted by $D^g$.}
  \label{fig:sp_illustration}
\end{figure}

We begin our investigation by examining the inherent characteristics of SPs in thermosetting networks. Fig. \ref{fig:sp_illustration}A and \ref{fig:sp_illustration}B illustrates two snapshots of the all-atom system with a SP highlighted. Initially, the SP appears slack, since its length is well beyond the system size. However, as deformation occurs, the SP gradually tightens and eventually fractures upon reaching a state of tautness. This observation aligns with identified global stress pathways in Fig. \ref{fig:local_stress}. Furthermore, Figure \ref{fig:sp_illustration}C presents the SP length distribution in systems with varying crosslinking degrees. Notably, as the network becomes more connected (highly crosslinked), the SP length distribution shifts towards shorter lengths, due to the increased availability of pathways. To analyze the evolution of the SP set under deformation, we present their contour lengths  relative to the system length (in the extended direction), as shown in Fig. \ref{fig:sp_illustration}D. The dashed line represents states where the contour length and the system length are equal, signifying a fully straightened SP. It can be seen that SPs tend to fracture upon nearing this limit. Exceptions occur where some SPs fracture prematurely, sharing edges with other SPs already stretched to near tautness. Additionally, the generally consistent slope of contour lengths vs. system size for the SPs in Fig. \ref{fig:sp_illustration}D implies that the shortest SP, noted as the first SP hereinafter, is more likely to first approach the dashed line, thereby indicating that fracture nucleation predominantly occurs on the first SP.  

\begin{figure}
  \centering
  \includegraphics[width=0.9\linewidth]{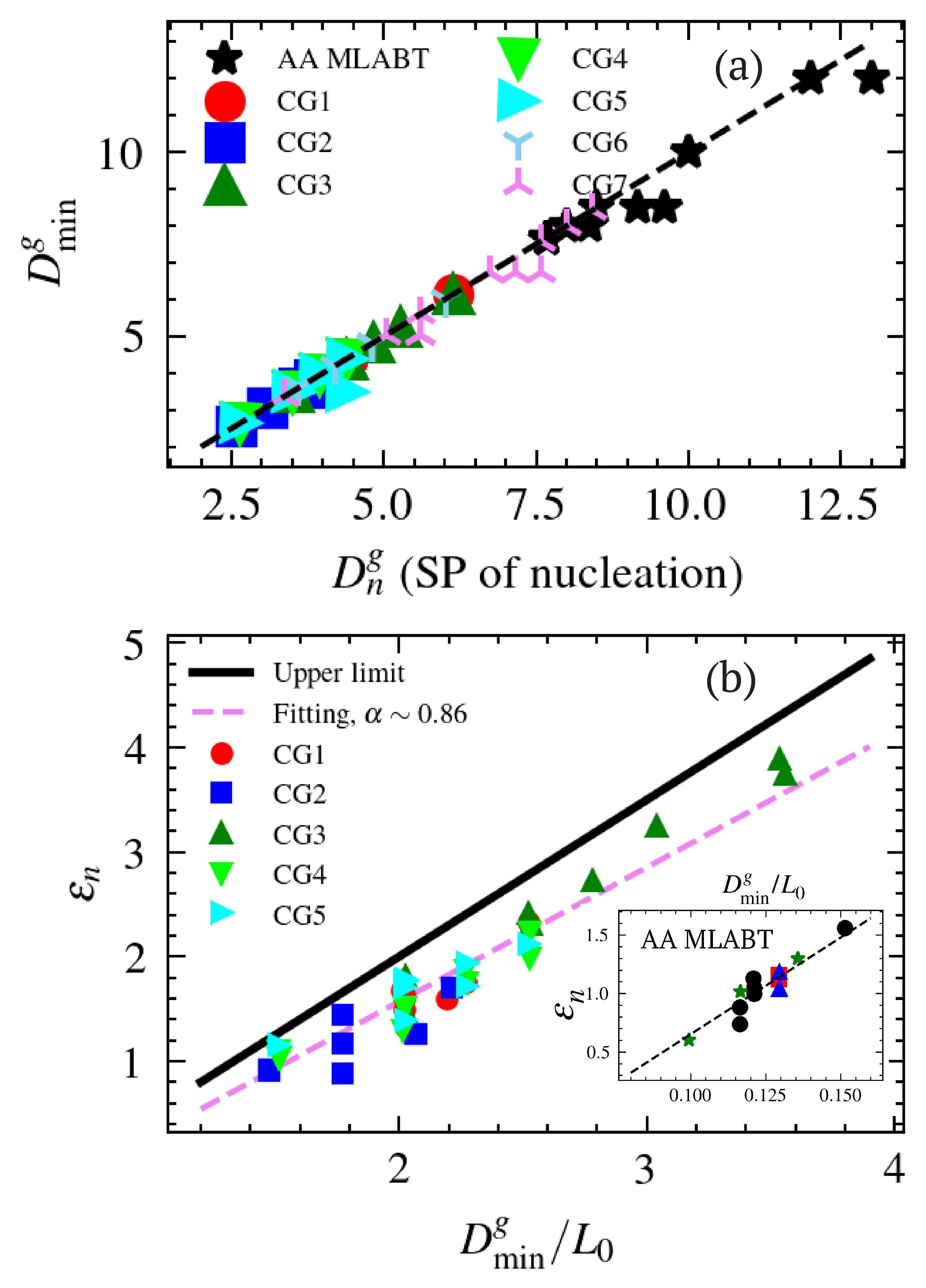}
  \caption{(a) Lengths of SPs on which fracture nucleation occurs, $D^g_{n}$, are consistent with minimum SP lengths $D^g_\mathrm{min}$ in both the all-atom and CG models under diverse conditions. The conditions examined are detailed in Table S1. (b) Linear correlation between the strain of fracture nucleation and the normalized minimum SP length in the CG models and the all-atom models (inset). The fits are based on Equation 2. }
  \label{fig:sp_nulceation}
\end{figure}

To statistically assess the prevalence of fracture nucleation occurring on the first SP, we investigate both the all-atom and the CG models under a variety of conditions. As shown in Fig. \ref{fig:sp_nulceation}A, the lengths of the SPs where fracture nucleation occurs, $D^g_n$, are generally consistent with the length of the first SP (or the minimum SP length, denoted as $D^g_\mathrm{min}$). The simulation conditions examined are provided in Table S1, covering variations in degrees of crosslinking, temperatures, strain rates, and strand lengths, reinforcing the hypothesis that fracture predominantly nucleates along the shortest SP. An ensuing question pertains to the precise nucleation location on a given shortest SP. To address this, we conduct 100 independent CGMD runs, incorporating cooling and deformation processes, on a network with a fixed topology.
These simulations revealed a non-uniform probability distribution for bond scission along the first SP, as detailed in Sec. S3. Typically, bonds whose rupture would result in the greatest stress relief are favored, as indicated by SP betweenness centrality or by the increase in $D^g_\mathrm{min}$ should that bond break, in agreement with recent computational studies.\cite{zhangPredictingFailureLocations2024} This finding underscores the possibility of predicting the fracture nucleation site at the molecular level through detailed SP analysis.

Beyond pinpointing the location of bond scission, the SP analysis also predicts the onset strain for fracture nucleation. As illustrated in Fig. \ref{fig:sp_illustration}B and \ref{fig:sp_illustration}D, the initial bond breakage is expected as the first SP straightens, i.e., its contour length approximately matches the system length in the elongated direction $L_n$, 
\begin{equation}
    L_n \simeq \alpha \cdot D^g_\mathrm{min} \cdot \bar{l} 
\end{equation}
where $\bar{l}$ is the average bond (or edge) length within the SP at the point of fracture nucleation and $0 < \alpha \leq 1$ is a geometric factor that accounts for premature breakage of the SP before complete straightness (e.g., due to inherent structural constraints). 
Given $L_n=(\varepsilon_n+1)L_0$ and $\bar{l} \simeq \lambda l_0$, where $L_0$ is the original system length, $l_0$ is the equilibrium bond length, and $\lambda$ the average bond stretching factor at breakage, the fracture nucleation strain $\varepsilon_n$ can be approximated as  
\begin{equation}\label{eq:fitting}
    \varepsilon_n \simeq \alpha \lambda l_0 \frac{D^g_\mathrm{min}}{L_0} - 1
\end{equation}
Hence, $\varepsilon_n$ is expected to have a linear relationship with $D^g_\mathrm{min}/L_0$, illustrating its dependence on a global topological characteristic. 

This fact is supported by a linear correlation between the normalized minimum SP length $D^g_\mathrm{min}/L_0$ and the fracture nucleation strain observed in both the all-atom models and CG models under a variety of simulation conditions, as shown in Fig. \ref{fig:sp_nulceation}B. By setting $\alpha=1$ (and $\lambda \sim$1.25 in the CG model), an upper limit of $\varepsilon_n \le \lambda l_0 D^g_\mathrm{min}/L_0 - 1$ can be derived, represented by the black line in Fig. \ref{fig:sp_nulceation}B. Notably, all data points from the CG models fall below this upper limit. Moreover, 
they can be described by a united fitting based on Equation \ref{eq:fitting}. With $\lambda$ and $l_0$ held constant, $\alpha$ is calculated to be 0.86, corresponding to a tilt angle of 17$^{\circ}$ for the first SP deviating from complete straightness at the moment of breakage. Based on this linear relationship, one can predict the fracture nucleation strain by computing the minimum SP length directly from the initial thermoset topology without resorting to MD simulations. 

\begin{figure}
  \centering
  \includegraphics[width=0.9\linewidth]{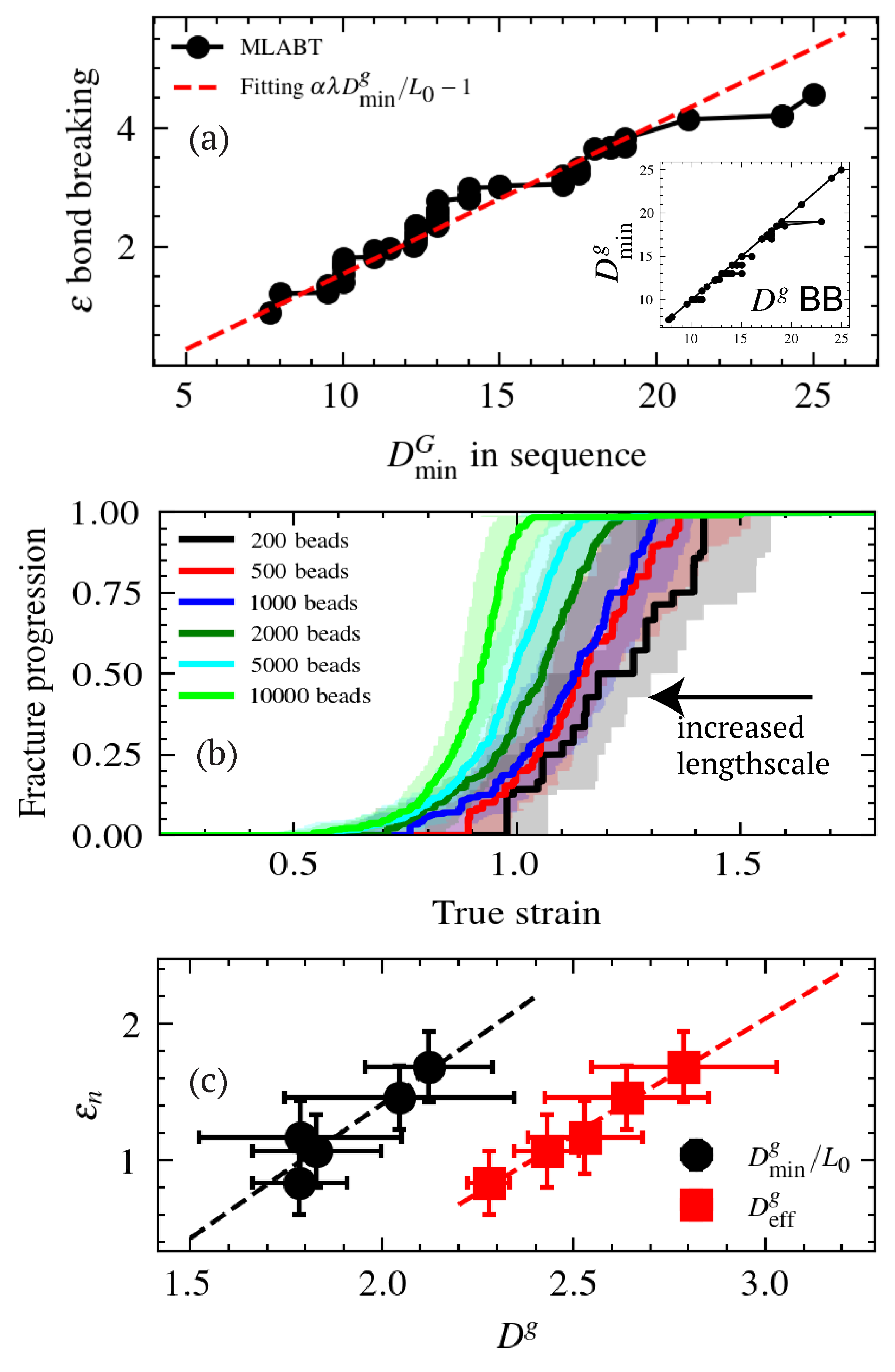}
  \caption{(a) Strains of sequential bond breaking events after nucleation  are approximately linearly correlated with the instantaneous minimum SP length in the all-atom model. The inset shows that lengths of SPs where the bonds break are consistent with the minimum SP lengths, suggesting that most bond breakages happen on the instantaneous first SP. (b) Fracture occurs at smaller strains when the system size is increased in the CG model. The fracture progression is defined as the number of broken bonds divided by the number of final broken bonds at complete network rupture. (c) Correlation of the averaged fracture nucleation strains with the averaged minimum SP length as well as the averaged effective SP length at various system sizes. The shadow regions in (b) and the error bars in (c) represent the standard deviations derived from at least 10 independent replicas. }
  \label{fig:SP_seq_scale}
\end{figure}

Fracture events following nucleation can also be analyzed through the lens of SPs in dynamically evolving networks. When the strain rate is sufficiently low to allow for stress redistribution, subsequent bond breakages behave similarly to initial nucleation events within the partially fractured networks. As shown in the inset of Fig. \ref{fig:SP_seq_scale}A, the lengths of SPs where the sequential bond breakages occur generally align with the length of the first SP in the updated network, suggesting a propensity for fracture to continue along the instantaneous first SP. Moreover, the established linear relationship between the strain at bond breakage and the minimum length of the instantaneous SP is adequately maintained in both the all-atom model (Fig. \ref{fig:SP_seq_scale}A) and the CG model (Fig. S6B). However, at higher strain rates (such as 10$^{11}$/s in the all-atom model), redundant bond breakages may occur along previously fractured SPs due to local stress concentration. This phenomenon becomes more pronounced in larger systems, as evidenced by the increased rate of bond breakage (Fig. \ref{fig:SP_seq_scale}B), because the timescale required for stress relaxation scales with the structural lengthscale. Thus, while SP analysis remains a powerful tool for understanding molecular-level fracture nucleation, at the mesoscale, crack propagation driven by localized stress concentrations becomes increasingly significant.\cite{longFractureHighlyDeformable2021,livneNearTipFieldsFast2010} Despite the complexity of fracture growth dynamics, the SP framework consistently provides predictive insights into fracture nucleation across scales and conditions. 

Importantly, we observe considerable finite size effects in fracture nucleation governed by SPs. As demonstrated in Fig. \ref{fig:SP_seq_scale}B, with increasing system sizes in the CG model, fracture occurs at smaller strains, a trend that includes fracture nucleation. This trend qualitatively agrees with that when varying sample sizes in experiments.\cite{fiedlerFailureBehaviorEpoxy2001}  More details, including the stress-strain curves, are provided in Sec. S5 of the Supplementary Material. The observed shift in nucleation strains implies a lengthscale dependency of SP characteristics. However, the minimum SP length tends to converge (also shown in the network model in Sec. S5D) after some initial decline and cannot explain the continual decrease of nucleation strains with increasing system size. The reason lies in larger systems hosting a greater number of SPs with lengths closely matching the minimum SP length, as indicated by narrower SP length distribution and a continuous decrease of average SP length toward the minimum length (Fig. S7). To account for the prevalence of nearly degenerate SP lengths in large systems, we introduce an effective SP length, $D^g_\mathrm{eff} = \frac{\sum_i 1}{\sum_i (L_0/D^g_i)}$, where the summation is operated over the SP set, see Sec. S5E. Despite the convergence of the minimum SP length, the effective SP length maintains a linear relationship with fracture nucleation strain across system sizes, as illustrated in Fig. \ref{fig:SP_seq_scale}C. This correlation is valid across all CG models evaluated, irrespective of the force fields applied (see Sec. S5). Therefore, the tendency for earlier fracture nucleation in larger systems can be attributed to a shift in the SP distribution towards shorter lengths. 

This lengthscale dependence of SPs sheds light on the significant discrepancy between brittle fractures observed in experiments (strain$<$0.1) and ductile fractures predicted in molecular simulations (typically strain$>$1).\cite{buckleyDeformationThermosettingResins2001,wuMultiscaleModelingEpoxies2020,mengPredictingMacroscopicFracture2016} This disparity can be at least partially addressed without the need to invoke defect structures. The SP picture suggests that expanding the size of amorphous networks, which comprise similar local structures, will naturally accelerate fracture nucleation. Moreover, in larger systems, fracture propagation is likely expedited as crack growth becomes increasingly significant. By extrapolating the effective SP length and following the linear relationship with nucleation strain, it may be possible to bridge molecular simulation results to experiments, despite the current absence of all-atom simulation data across varying lengths. 

In summary, we introduce a definition of the SP concept within polymer networks, establishing its direct correlation with fracture nucleation. This approach allows us to predict the location and strain at fracture nucleation, using only the network's initial topology. Furthermore, SP analysis offers a coherent framework for interpreting previous simulation results.\cite{yuExploringThermosetFracture2024} Temperature, thermal fluctuations, and cooling rates have minimal impact on fracture behaviors since they are unrelated to network topology and SP. Conversely, the degree of crosslinking significantly influences fracture dynamics by altering the SP length distribution (see Fig. \ref{fig:sp_illustration}B). Although strain rate does not affect the initial topology, it alters the subsequent network topology. At high strain rates, additional bonds break along previously broken SPs, increasing bond breakage rates but not affecting the strain at fracture nucleation.  

The analysis of lengthscale dependence through SPs holds additional promise for bridging the gap between experimental observations and simulation predictions of thermoset fractures. By modifying the network formation process, chemical composition, and polymerization degree, we can substantially alter SP characteristics and, consequently, the fracture behaviors of polymer networks. This SP-centric analysis not only sheds light on the microscopic mechanisms underlying fracture in polymer networks but also opens new pathways for designing thermosets with customized fracture properties, signifying a shift towards more predictive and tailored material engineering.

\begin{acknowledgments}
This material is based upon work supported by the National Science Foundation Chemical Theory, Models, and Computation division under award CHE-2154916. This work used Bridges-2 at the Pittsburgh Supercomputing Center through allocation CHE230055 from the Advanced Cyberinfrastructure Coordination Ecosystem: Services \& Support (ACCESS) program, which is supported by National Science Foundation grants \#2138259, \#2138286, \#2138307, \#2137603, and \#2138296.
\end{acknowledgments}

\bibliography{SP}

\end{document}


\title[]{Supplementary Material for ``Shortest paths govern fracture nucleation in thermoset networks''}

\author{Zheng Yu}
\affiliation{Department of Chemistry, University of Illinois at Urbana-Champaign, Urbana, Illinois, 61801, USA}

\author{Nick Jackson}%
 \email{jacksonn@illinois.edu}
\affiliation{Department of Chemistry, University of Illinois at Urbana-Champaign, Urbana, Illinois, 61801, USA}


\maketitle

\tableofcontents
\section{Methods}\label{sec:M}

\subsection{Searching shortest paths in networks with periodic boundary conditions}

In a graph $G(V,E)$, where $V$ denotes nodes and $E$ denotes edges, a path is defined as any sequence of nodes such that every consecutive pair of vertices in the sequence is connected by an edge in the network.\cite{newmanNetworks2018} The length of a path is the number (or sum of weights) of edges along the path. A shortest path (SP), also called a geodesic path, is a path between two nodes such that no shorter path exists. The length of a SP, i.e., geodesic distance, between nodes $u_i$ and $u_j$ in an unweighted network can be computed by $D^g(u_i,u_j) = \mathrm{argmax}_r [A^r]_{V_i,V_j}>0$. In large sparse networks with non-negative weighted edges, geodesic distances between any nodes can be efficiently computed using Dijkstra's algorithm in time complexity of $\mathcal{O}(n\log n)$.

\begin{figure}[H]
  \centering
  \includegraphics[width=0.9\linewidth]{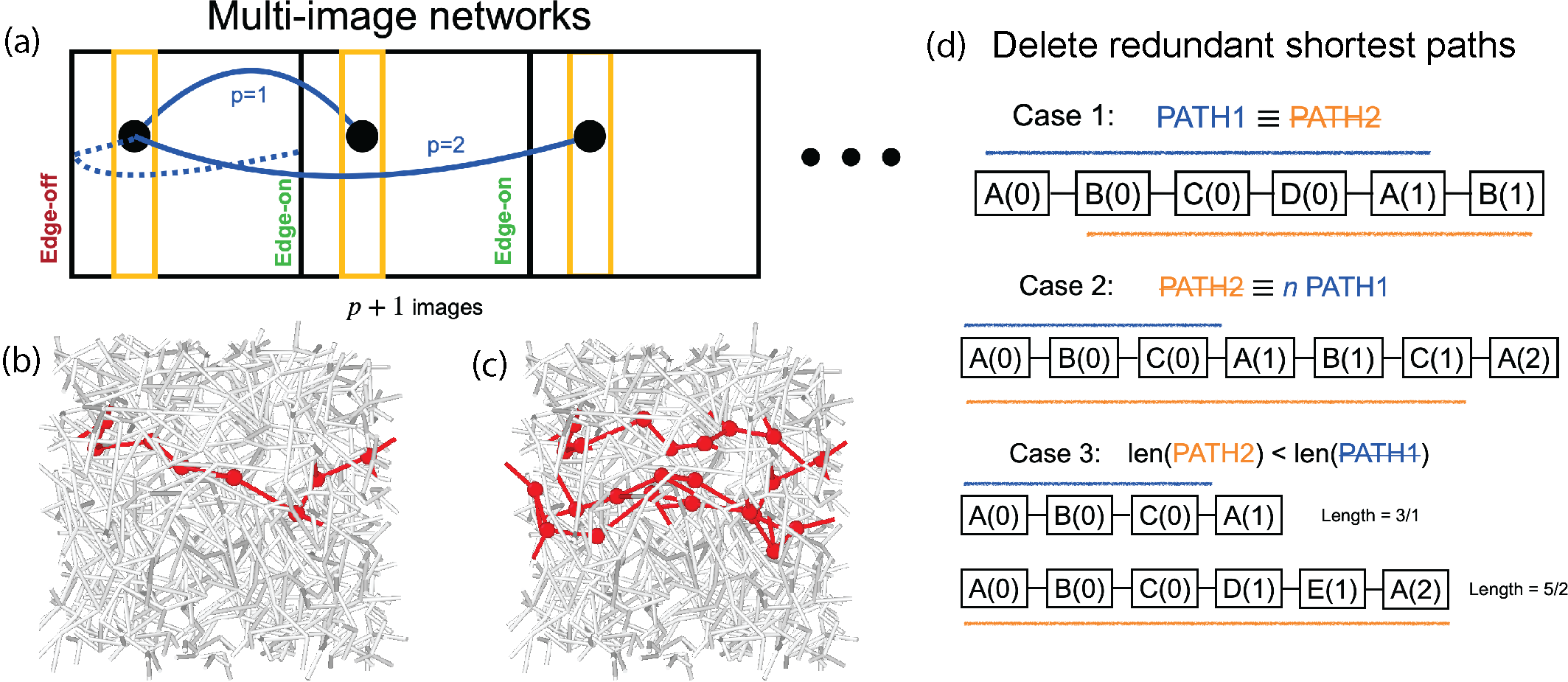}
  \caption{(a) Illustration of shortest paths involving multiple periodic images. (b) An example of a shortest path with $p=1$ from the all-atom model. (c) An example of a shortest path with $p=3$ from the all-atom model. Note that the three path segments in one image are connected to each other at periodic boundaries. (d) Three cases of redundant shortest paths that need to be deleted after searching.}
  \label{fig:multiple_wavelength}
\end{figure}

The key question is how to define $G$ given the material system and select node pairs between which the SP will potentially sustain maximum stress under deformation. Intuitively, the nodes $\{u_i | i = 1,2...,N\}$ represent the particles (atoms, molecules, coarse-grained beads, etc.), where $N=|u|$ denotes the total number of particles, and the edges represent the direct covalent bonds or a sequence of covalent bonds without intermediary nodes. Considering the bulk network materials with the periodic boundary condition (PBC) in the simulation context, three steps are designed to completely search all shortest paths without human intervention. 

i. Create a series of multi-image networks, $G_1$,$G_2$,..., where $G_p$ is a network containing $p+1$ periodic images. This consideration is made because SPs in periodic systems can have wavelengths of more than one image, as illustrated in Fig. \ref{fig:multiple_wavelength}. For each multi-image graph, edges can be formed across the inner periodic boundaries but not across the outer boundaries to ensure all SPs pass through all the images. 

ii. Select pairs of identical particles $u_i$ from the farthest images and compute the SP using Dijkstra's algorithm $SP(u_i^0, u_i^p; G_p)$, where $u_i^p$ denotes the particle $u_i$ in the $p$th periodic image. By iterating over $p$, a sufficiently large number of images, and over $i$, all particles in the system, we can obtain a complete set of SPs, $\{SP(u_i^0, u_i^p; G_p) | p = 1,2,...,M; i = 1,2,...N \}$, where $M$ is a large number determined by testing in practice. 

The length of SP from multi-image networks is normalized by the number of images, 
\begin{equation}
    D^g(u_i^0, u_i^p; G_p) = \frac{1}{p} \text{ShortestPathLength}(u_i^0, u_i^p; G_p)
\end{equation}
where ShortestPathLength denotes the reported length of SP directly from the Dijkstra's algorithm. This normalization step is particularly important because the length needs to be compared with the system size for prediction. 

iii. Delete redundant shortest paths. There are three cases that might be deleted, as illustrated in Fig. \ref{fig:multiple_wavelength}d. 1) Identical SPs that start at different nodes. 2) SPs with larger $p$ but repeat those with smaller $p$. This will result in the convergence of searching at a finite $p$ value. 3) Among SPs between identical nodes $\{\mathrm{SP}(u_i^0,u_i^p;G_p) | p = 1,2,...,M\}$, only the one with the shortest length is remained and others are deleted.  

In practice, a relatively large $M$ (e.g., 6-8 in our models) is sufficient for $p$, because in networks with too large $p$, there are no more unique SPs and all SPs are repeated patterns of those with smaller $p$. In addition, we might select a slice parallel to the direction of applied strain and only search for nodes in the slice. This will reduce the computational cost, especially when the system is large. 

\subsection{Computational models of thermoset networks}\label{sec:computational_models}

\begin{figure}[H]
  \centering
  \includegraphics[width=0.9\linewidth]{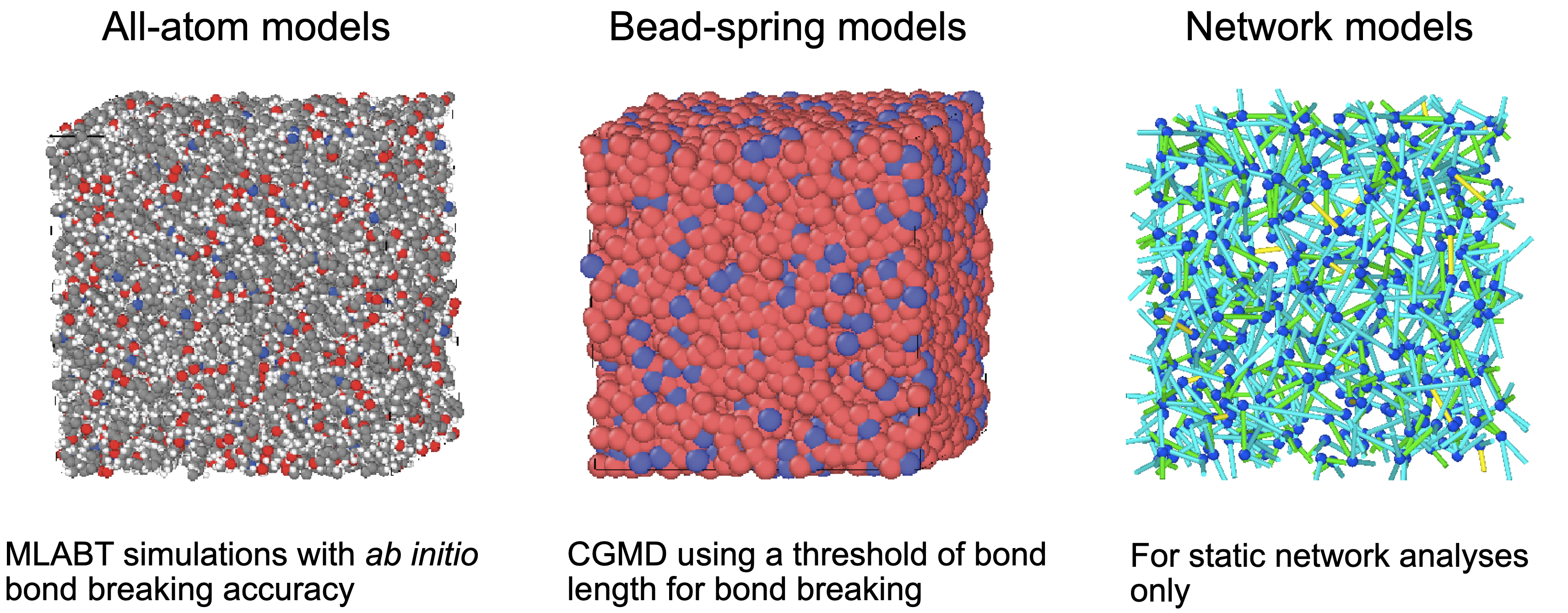}
  \caption{Three computational models are investigated in this work.}
  \label{fig:computational_models}
\end{figure}

In this work, three different models have been investigated to understand the comprehensive features  of the SP mechanism. The three models are illustrated in Fig. \ref{fig:computational_models} and detailed as follows.

\textbf{All-atom models}. We employ an archetypal epoxy polymer system, diglycidyl ether of bisphenol A (DGEBA) cured by methylene dianiline (MDA), as investigated in our previous studies.\cite{yuMachineLearningQuantumchemical2023} We apply the machine learning-based adaptable bonding topology (MLABT) method for simulating the molecular fracture process, a machine learning (ML) method incorporated on-the-fly with classical MD to accurately describe quantum-chemically accurate bond breaking at dramatically reduced cost. The ML model has been trained with an active learning strategy using data collected from density functional theory (DFT) calculations.\cite{yuExploringThermosetFracture2024,neeseORCAQuantumChemistry2020}

Specifically, the all-atom model contains 432 DGEBA and 216 MDA molecules (27,432 atoms in total) in a cubic box with periodic boundary conditions in three dimensions. Bonding topologies of networks are generated dynamically by simulating curing reactions in MD, resulting in degrees of cross-linking ranging from 77\% to 98\%. Structures are melted at 800 K for 200 ps and then quenched to 300 K with a constant cooling rate ranging from 0.1 to 100 K/ps. The obtained glassy structures are then used as the initial conditions for MLABT deformation simulations. Only uniaxial deformations are considered in this work. During deformations, the simulation box is deformed uniaxially (e.g., along the x-axis) every 0.025 ps at a strain rate of 1 × 10$^9$/s, and the atomic coordinates are remapped accordingly. The two transverse directions are allowed to relax under P = 1 atm to avoid the accumulation of artificial stress. We apply the optimized potentials for liquid simulations all atoms (OPLS-AA) force-field with the large-scale atomic/molecular massively parallel simulator (LAMMPS) in all MD simulations.\cite{jorgensenDevelopmentTestingOPLS1996,LAMMPS} The simulations are carried out on the Bridges-2 cluster, which is provided by the Advanced Cyberinfrastructure Coordination Ecosystem: Services \& Support (ACCESS).\cite{boerner2023access}

\textbf{Coarse-grained models}. The non-bonded interaction is set as the Lennard-Jones interaction, i.e., $V_\mathrm{LJ} = 4 \epsilon [(\frac{\sigma}{r})^{12}-(\frac{\sigma}{r})^6]$, where $\epsilon=1$ and $\sigma=1$, and the cutoff distance is 2.5. Two types of bonded interactions are tested, separately. One is quartic bonds,\cite{stevensInterfacialFractureHighly2001}
\begin{equation}
    V_q = K(r-R_c)^2(r-R_c-B_1)(r-R_c-B_2)+V_0
\end{equation}
where $K=1200$, $B_1=-0.55$, $B_2=0.25$, $R_c=1.3$, and $V_0=34.6878$ in order to mimic the finite extensible nonlinear elastic (FENE) potential.\cite{kremerDynamicsEntangledLinear1990} When the bond length is above $R_c$, there is no interaction between the two associated beads and the bond is effectively broken. We also apply the harmonic bonds (as applied in the all-atom models), i.e., $V_h=K(r-R_0)^2$, where $K=100$ and $R_0=0.8$. A cubic box with PBC in the three dimensions is used, containing beads with numbers ranging from 50 to 1 million. Two types of beads are used, A as monomers and B as crosslinkers. For simplicity, A and B are simulated with the same force field parameters. The polymer chains formed by A have lengths of 2/4/6 beads in different systems. The functionality of B is set to 3 and that of A is 2. 

The CG network is formed dynamically, similar to that in the all-atom models. Below list the specific steps employed for generating stable network structures. 
\begin{itemize}
    \item Prepare initial polymer chains and crosslinkers in a simulation box. For simplicity, the polymer chains are generated on lattice with the crosslinkers randomly located around them, ensuring that no beads are located within 0.8 of another bead to prevent simulation instability. 
    \item Relax initial structures in the isobaric-isothermal ensemble (NPT) at high temperatures. This helps to prepare amorphous structures for the subsequent curing step. 
    \item Link polymers and crosslinkers by simulating the curing reactions. This step is done during annealing simulations at high temperatures. Once a crosslinker and a polymer bead get close to each other within some cutoff distance, a bond is immediately formed between the two and the simulation is resumed. Note that if the degree of a bead reaches its functionality, the bead cannot participate in reactions anymore. This step stops when the desired degree of crosslinking is achieved, which is defined as the number of formed bonds divided by the number of crosslinkers times the functionality.  
    \item Melt and then quench the system to obtain stable structures. This is a routine step for glass simulations. It usually requires a relatively small cooling rate (1e-5/timestep is used) but actually the cooling rate has been reported not critical to fracture mechanics.\cite{yuExploringThermosetFracture2024} 
    \item Perform deformation simulations on the structures. The simulation box is stretched uni-axially (e.g., the $x$ direction) at a constant true strain rate (0.005/timestep is used unless otherwise stated), and the sizes of the other two dimensions are kept fixed.  
\end{itemize} 

All the simulations are performed with LAMMPS and a constant timestep of 0.002 is used.\cite{LAMMPS} In addition, we find that the cutoff distance used for the crosslinking reactions significantly influences the generated topology and corresponding fracture properties. For instance, a too large cutoff, such as twice of the equilibrium bond length, may lead to concentrated defects and early fracture nucleation, which can still be captured by the effective SP analyses. 

\textbf{Network models.} We also employ 3D network models just for SP analysis. To build the model, first nodes are randomly distributed in a cubic box with PBC in all directions. Second, a structure relaxation with the Morse potential is applied to make sure the node distribution is reasonable compared to those in physical systems. Finally, edges are formed between nodes based on a Gaussian probability, $P_{u_i,u_j}(r_{ij}) = \exp(\frac{1}{2}(r_{ij}/\sigma)^2)/\sum_{k<l} \exp(\frac{1}{2}(r_{kl}/\sigma)^2)$, where $r_{ij}$ is the distance between nodes $u_i$ and $u_j$, and $P_{u_i,u_j}$ is the associated probability of forming an edge. $\sigma$ is the standard deviation depending on the system, which could be related to the chain length if an ideal chain behavior is assumed. 

To mimic the all-atom models, we apply the same system settings as in the all-atom models, including the system size, the density of crosslinkers, the degree of crosslinking, and the functionality. However, because the crosslinking probability for pair distances is not exactly Gaussian in thermosets, the network models should not be considered an exact surrogate model of the all-atom models, but some generalized random physical networks. 

\begin{table}[H]
\caption{Details of the computational models for analysis in Fig. 3 in the main text. }\label{tab:table2}
\begin{ruledtabular}
\begin{tabular}{lccccc} 
& Number of particles & degrees of crosslinking & Temperature & Strand length & strain rate \\
\colrule
CG1 & 200 beads & 0.9  & 0.1 & 2 beads & 0.005 \\
CG2 & 1000 beads & 0.9  & 0.1 & 2 beads & 0.005 \\
CG3 & 200 beads & 0.75  & 0.1 & 2 beads & 0.005 \\
CG4 & 200 beads & 0.9 & 0.3 & 2 beads & 0.005 \\
CG5 & 200 beads & 0.9  & 0.1 & 2 beads & 0.001 \\
CG6 & 360 beads & 0.9 & 0.1 & 4 beads & 0.005 \\
CG7 & 520 beads & 0.9 & 0.1 & 6 beads & 0.005 \\
All-atom & 27432 atoms (432 crosslinkers) & 0.77-0.98 & 100-500 K & 1 monomer & 1e9-1e11/s
\end{tabular}
\end{ruledtabular}
\end{table}

\section{Shortest paths in the coarse-grained model}

\begin{figure}[H]
  \centering
  \includegraphics[width=1\linewidth]{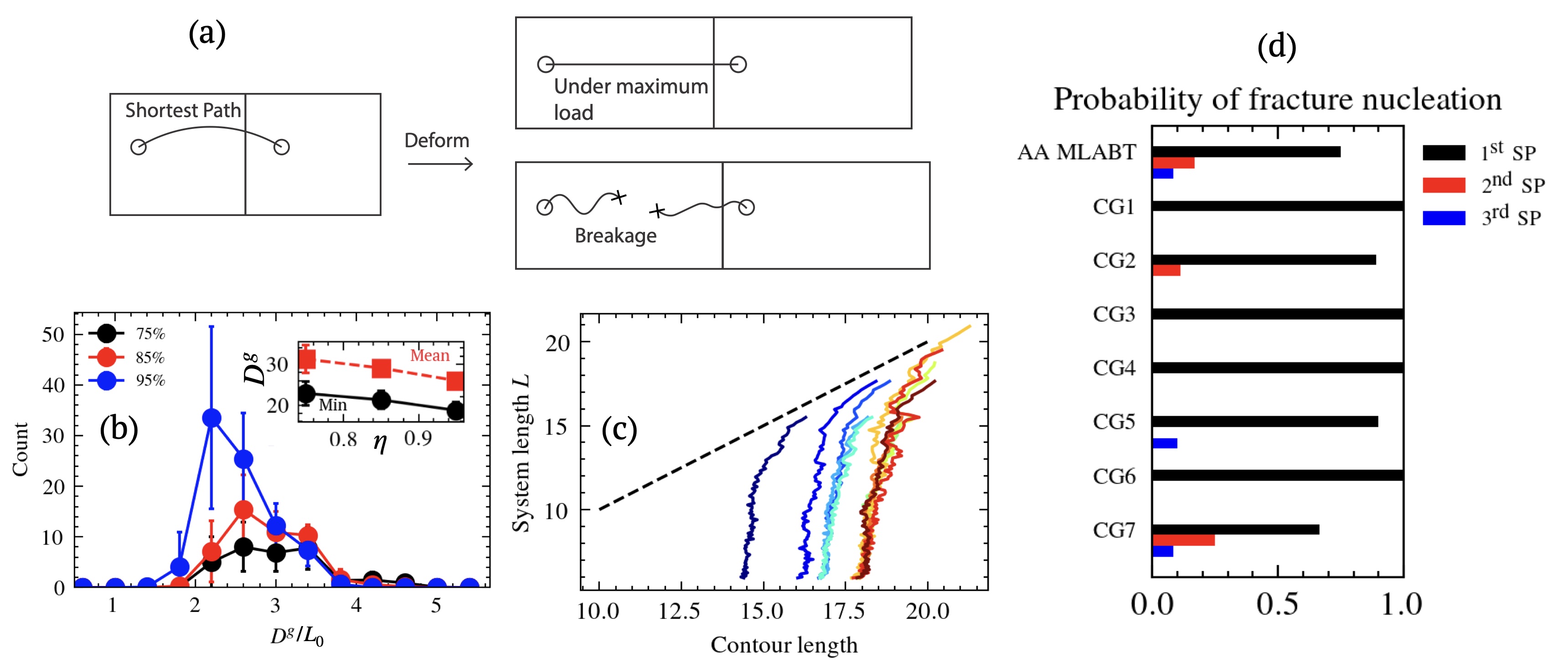}
  \caption{(a) Illustration of the SP scenario, which is consistent with the observation in the all-atom model in Fig. 2A in the main text. (b) Distributions of SP lengths in the CG models (harmonic bonds) at three degrees of crosslinking $\eta$=75\%, 85\%, and 95\%. The inset shows the corresponding change in the minimum and average length of SPs. (c) length evolution of SPs under extension compared to the system length $L$ in the elongated direction. The results are similar to the same figure for the all-atom model in Fig. 2C of the main text. (d) Probability of fracture nucleation on the first, second, and third SP in different models. }
  \label{fig:SP_nature_CG}
\end{figure}

As illustrated in Fig. \ref{fig:SP_nature_CG}A, SPs provide a clear picture of how fracture nucleation occurs under the constraint of periodic boundaries. Due to the relaxation capability of structures under deformation, SPs are expected to approach very close to straightness at breakage, similar to stretching a polymer chain. Therefore, the lengths of SPs are considered important features, and they vary with degrees of crosslinking. Similar to that in the all-atom model (Fig. 2 in the main text), at more connected networks, the SP length distribution tends to shift towards the shorter lengths (Fig. \ref{fig:SP_nature_CG}B), also indicated by the decreasing minimum and mean values in the inset. As depicted in Fig. \ref{fig:SP_nature_CG}C, different SPs share similar length evolution behavior under deformation, suggesting that the first SP (the shortest one) will be the first to approach tautness (indicated by the dashed line) and break. This is consistent with the observation in the all-atom model presented in Fig. 2D of the main text. Moreover, Fig. 3A of the main text clearly shows that fracture nucleation mostly occurs on the first SP. To reinforce this argument, we present the probability of fracture nucleation occurring on the first three SPs in Fig. \ref{fig:SP_nature_CG}D. The conditions of the models are detailed in Table \ref{tab:table2}. It can be clearly seen that the first SP is more probable to nucleate the first bond breakage. Regarding the exceptions, there seems no universal explanation, but some might be related to bonds shared among multiple near-shortest SPs or bonds under strong local structural constraints, which warrants further investigation. 

\section{Specific bond breaking sites on the shortest path}

\begin{figure}[H]
  \centering
  \includegraphics[width=0.6\linewidth]{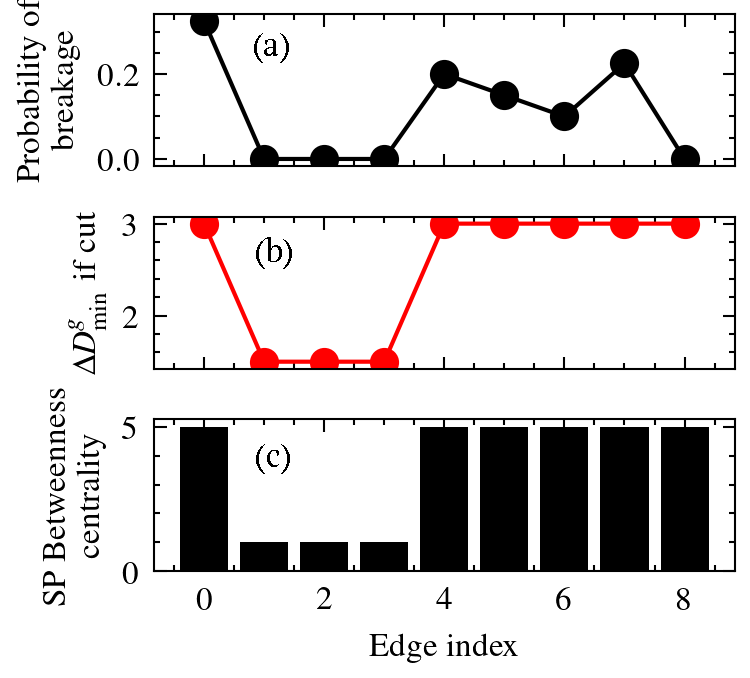}
  \caption{Investigation of the exact bond breaking sites on the first SP in a CG model with a fixed network topology. (a) Probability of breaking for each bond on the first SP. This is done by cooling and deformation 100 times with different initial velocities. (b) Increase of minimum SP length if a particular bond is cut, which are computed from all possible residual networks. (c) SP betweenness centrality of the bond, i.e., the number of SPs passing through the bond. }
  \label{fig:first_BB_location}
\end{figure}

Although we learn that the fracture nucleation mostly occurs on the 1st SP, exactly where on the path remains a question. To tackle this, we perform statistical analysis by running 100 cooling and deformation simulations on fixed-topology networks, in which the influence of glassy structures and thermal fluctuation will be reflected in the distribution of the bond breaking locations. As shown in Fig. \ref{fig:first_BB_location}A as one example, not every bond has equal probability to break, suggesting that breaking along the SP is different from breaking along a simple polymer chain. We have two theoretical hypotheses to understand this. One is the increase of minimum SP length, $D^g_\mathrm{min}$, if one particular bond on the SP is deliberately cut, as shown in Fig. \ref{fig:first_BB_location}B. Since $D^g_\mathrm{min}$ for the residual network roughly predicts the strain of next bond breakage (Fig. 4A), cutting the bond with the largest $D^g_\mathrm{min}$ increase tends to maximize the delay of the next bond breakage, in other words, to maximize the release of current stress. 

The other hypothesis concerns the SP betweenness centrality, i.e., the number of SPs passing through a particular bond. Betweenness centrality is a critical measure in network analysis, especially for flow dynamics and network disruption. In the problem of thermoset fracture, since stress is more likely distributed on the SP (especially the shorter ones), counting the SP betweenness centrality will shed insight on the most vulnerable bonds, i.e., bonds under the most loads, and naturally cutting those bonds will release more stress than cutting others. As shown in Fig. \ref{fig:first_BB_location}C, the results are similar to the $D^g_\mathrm{min}$ increase as in Fig. \ref{fig:first_BB_location}B. They both suggest that bonds with index 0 and 4-8 will have higher probability to break than bonds with index 1-3. But also note that the probability in \ref{fig:first_BB_location}A shows fluctuation, especially for bond index 0 and 8, not exactly following the two hypotheses, suggesting that there might be additional influence from local topology as well. In addition, computing the SP betweenness centrality is more efficient than computing the $D^g_\mathrm{min}$ increase, since only one SP search is needed. 

Therefore, while site of fracture nucleation is not deterministic nor entirely random, the probability for it to happen on individual bonds can be predicted from the SP analyses as discussed above.

\section{Fracture events after nucleation}

\begin{figure}[H]
  \centering
  \includegraphics[width=0.5\linewidth]{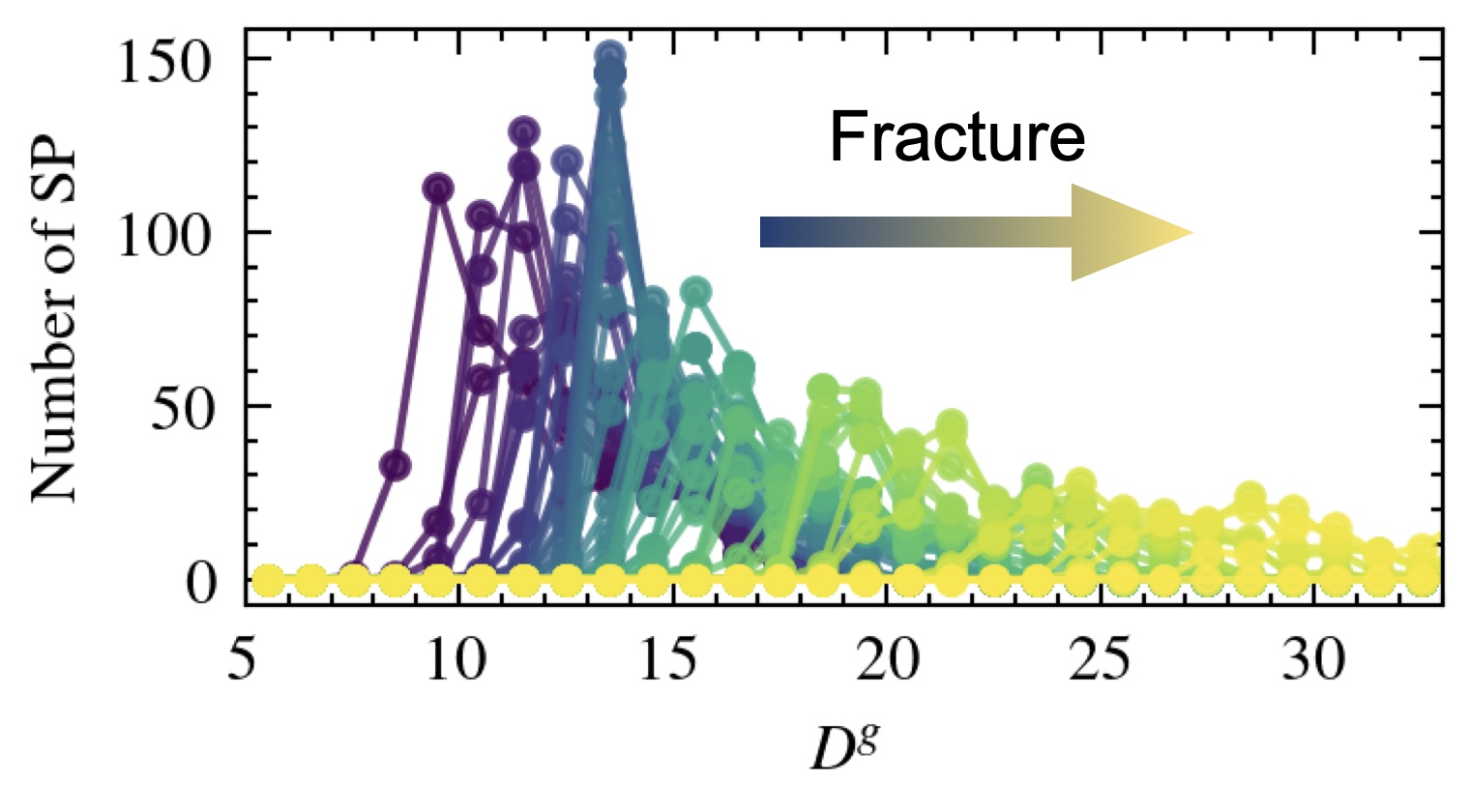}
  \caption{Evolution of SP length distribution during the MLABT fracture simulation in the all-atom model. As the network becomes more broken, the SP lengths increase and the numbers of SPs decrease. }
  \label{fig:sequence_AA_dist}
\end{figure}

As the network fractures, the connectivity diminishes and the SP lengths generally increase, as shown in Fig. \ref{fig:sequence_AA_dist}. This aligns with the observations when varying the degree of crosslinking. It also can be seen that the minimum SP length continuously increases during the fracture, suggesting again that the fracture events are related to the minimum SP length. 

\begin{figure}[H]
  \centering
  \includegraphics[width=0.9\linewidth]{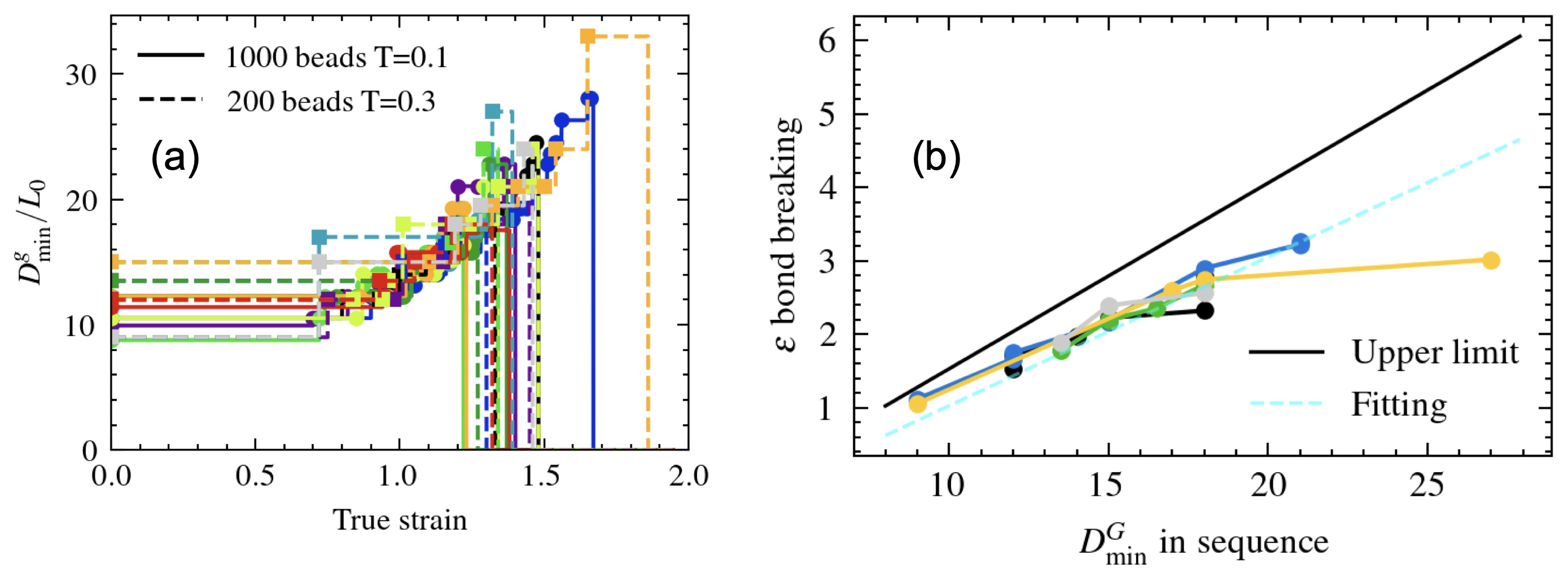}
  \caption{Instantaneous network SP analyses during deformation in the CG model. (a) Normalized minimum SP length as a function of true strain. The solid lines (with circles) are results of the model with 1000 beads at a temperature of 0.1, and the dashed lines (with squares) are those with 200 beads at temperature of 0.3. Colors represent different individual replicas. It can be seen that the evolution of instantaneous SP length is generally consistent, especially for the relatively large model. (b) The strain for the next bond breaking is approximately linear in the instantaneous minimum length of SPs, consistent with the observation in the all-atom model shown in Fig. 4A in the main text. }
  \label{fig:sequence_CG}
\end{figure}

We then compute the minimum SP length for each instantaneous network during deformation as a function of strain, as shown in Fig. \ref{fig:sequence_CG}A. The presented results include two different model conditions: 1000-bead systems deformed at a temperature of 0.1 and 200-bead systems deformed at a temperature of 0.3, with each condition containing 10 independent runs with different random initializations. The minimal SP lengths increase at each fracture event and eventually vanish at the failure point. It can be seen clearly that, despite the different conditions and initial topologies, the increasing trend of the minimum SP length generally follows the same behavior. This suggests that the subsequent fracture events after nucleation are still correlated with SPs. 

To evaluate this argument more clearly, we plot the strain at each bond breakage versus the minimum SP length of the instantaneous network before this breakage in Fig. \ref{fig:sequence_CG}B, similar to Fig. 4A but for the CG models. For better visualization, we only present a few cases in the figure and others not shown have consistent behaviors. Notably, a linear relationship is observed, which can be fitted based on Equation 2 in the main text. We also note that when the strain rate is too large, bond breakages would occur at the previously broken SPs due to limited stress relaxation. This effect will also lead to fracture mechanism transition to the crack propagation as the system size approaches the mesoscale. 

\section{Lengthscale dependence of shortest path lengths}

In the main text, we discussed how the SP length distribution depends on the system size (lengthscale) and consequently influences the fracture nucleation in the molecular simulation models. Because of the computational difficulty to obtain results of the all-atom model at vary large lengthscales, we focus on three CG models and a network model in this section, aiming to present a clear picture of the lengthscale dependence. 

\subsection{CG models with less stiff harmonic bonds (longer bonds, presented in the main text)}\label{sec:CGmodel_lh}

\begin{figure}[H]
  \centering
  \includegraphics[width=0.8\linewidth]{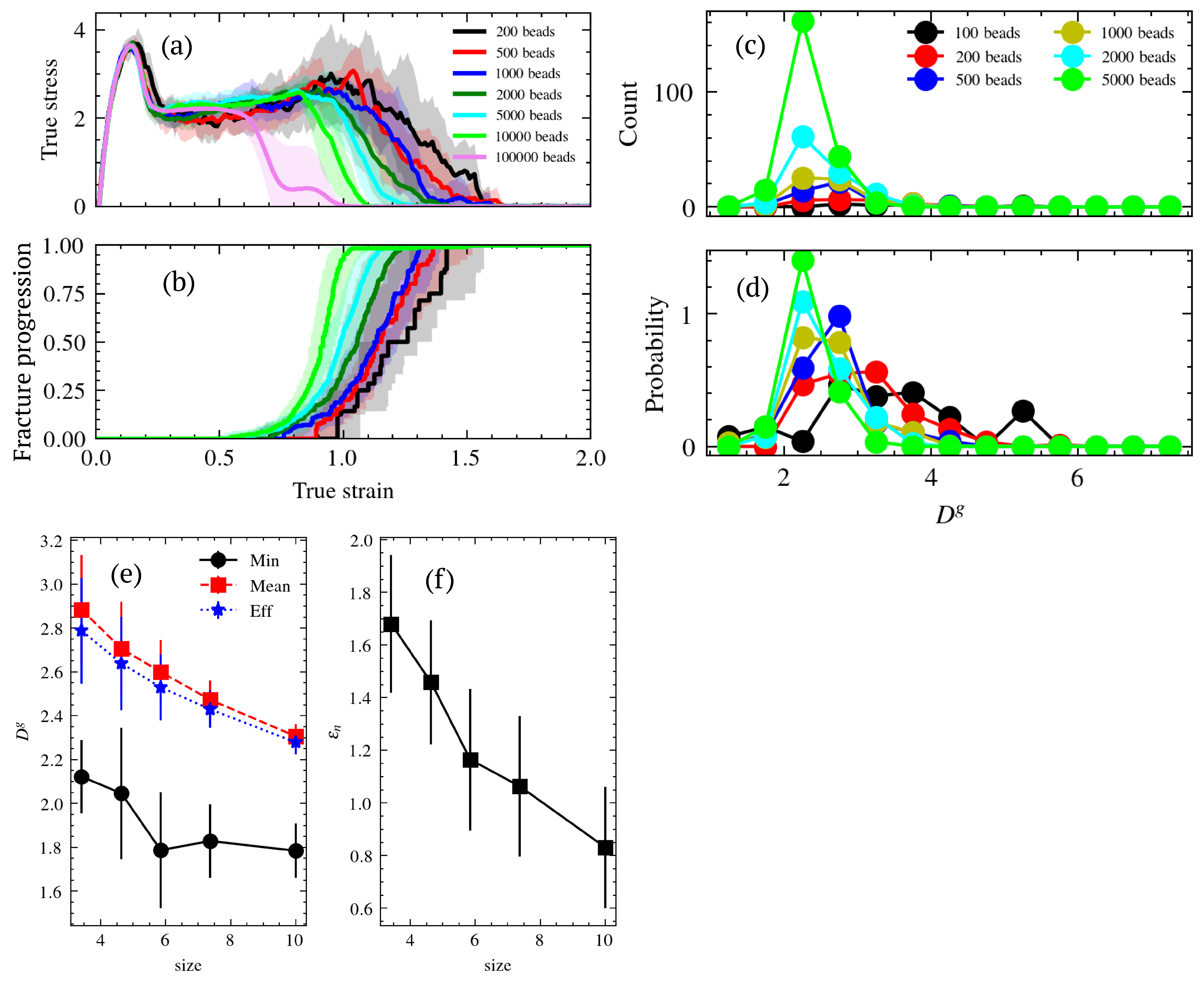}
  \caption{Dependence of fracture and SP lengths on the lengthscale. The parameters of the harmonic bonds in the CG model are the equilibrium bond length $l_0=1.2$, the bond strength $K=100$, and the breakage length criterion $l_c=1.5$. (a) Stress-strain curves of the CG model with harmonic bonds at various system sizes. (b) The fracture progression, which is defined as the number of broken bonds divided by the total number of broken bonds at complete failure, as a function of true strain. It is clear that the fracture is more brittle, i.e., occurring earlier, in larger systems. (c) Number and (d) density distribution of SP lengths in systems with various sizes. It is clear that the distribution becomes narrower in larger systems. (e) Minimum, mean, and effective length of SPs when the length is increased.  (f) Strain at fracture nucleation continuously decreases as the system becomes larger. } 
  \label{fig:SP_CG_harmonic_long}
\end{figure}

The first model we studied is the one presented in the main text Fig. 4. It employs harmonic bonds with the equilibrium bond length $l_0=1.2$, the bond strength $K=100$, and the breakage length criterion $l_c=1.5$. Details on generating the CG network can be found in Sec. \ref{sec:computational_models}. It can be seen clearly in Fig. \ref{fig:SP_CG_harmonic_long}A and \ref{fig:SP_CG_harmonic_long}B, with increasing system sizes and all other parameters kept fixed, the fracture of the CG model occurs at smaller strains, while the elastic behaviors are minimally affected. This suggests that the lengthscale effect is more pronounced on fracture than on elastic properties, in agreement with previous MD studies.\cite{zhaoUncoveringMechanismSize2022} The fracture nucleation strains decrease with increasing size, as shown in Fig. \ref{fig:SP_CG_harmonic_long}f. 

This size effect can be understood from the SPs. Figure \ref{fig:SP_CG_harmonic_long}C and \ref{fig:SP_CG_harmonic_long}D show the SP length distributions at varying system sizes. Even after normalization by the SP number, the distribution tends to narrow towards smaller lengths, showing a sharper asymmetric peak in larger systems. This suggests that in larger systems, there is an increased fraction of SPs having lengths close to that of the shortest SP, and this may expedite fracture nucleation statistically. To account for this effect, we introduce the effective SP length computed from the whole SP set, $D^g_\mathrm{eff} = \frac{\sum_i 1}{\sum_i (L_0/D_i)}$, where $L_0$ is the original system length and $D_i$ is the length of each SP. This form is similar to the resistance of parallel circuits but the numerator is $\sum_i 1$ instead of 1 to incorporate the change in the number of SPs (also resulting in the effective length larger than the normalized minimum length $D^g_\mathrm{min}/L_0$). An explanation of this form is provided in Sec. \ref{sec:effective_length}. Figure \ref{fig:SP_CG_harmonic_long}E presents the minimum, mean, and effective lengths of SPs as a function of system sizes. The minimum SP length decreases at first but converges around a size of 6, corresponding to the kink position in Fig. \ref{fig:SP_CG_harmonic_long}F. Instead, the effective and mean lengths are very similar and decrease continuously with increasing system sizes. Figure 4C in the main text shows clearly that the effective length has a better linear correlation with the fracture nucleation strain than the minimum length across varying system sizes. This is also true in all the other models we investigated. 

\subsection{CG models with less stiff harmonic bonds (shorter bonds)}

\begin{figure}[H]
  \centering
  \includegraphics[width=0.5\linewidth]{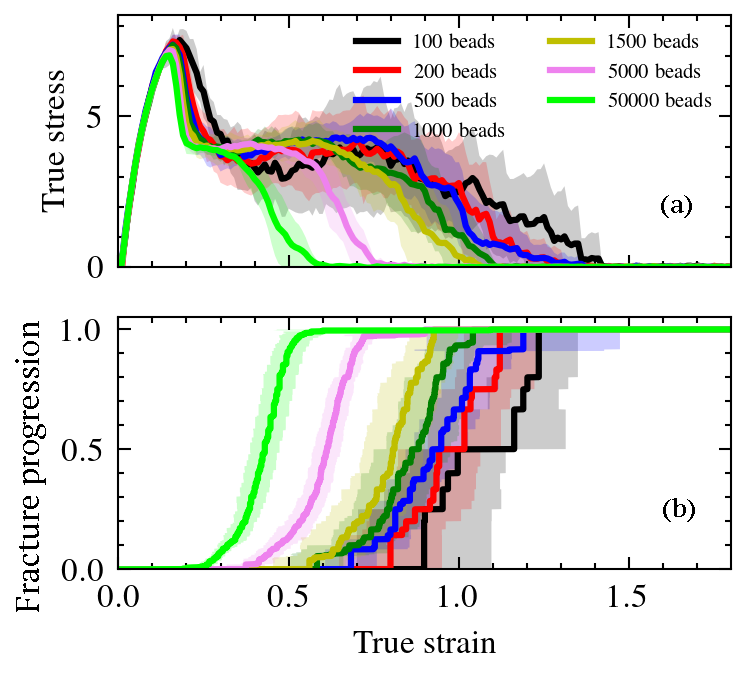}
  \caption{(a) Stress-strain curves of the CG model with harmonic bonds with increasing system sizes. (b) The fracture progression as a function of true strain. The parameters of the harmonic bonds in the CG model are the equilibrium bond length $l_0=0.8$, the bond strength $K=100$, and the breakage length criterion $l_c=1.1$.}
  \label{fig:deform_CG_sb}
\end{figure}

\begin{figure}[H]
  \centering
  \includegraphics[width=0.9\linewidth]{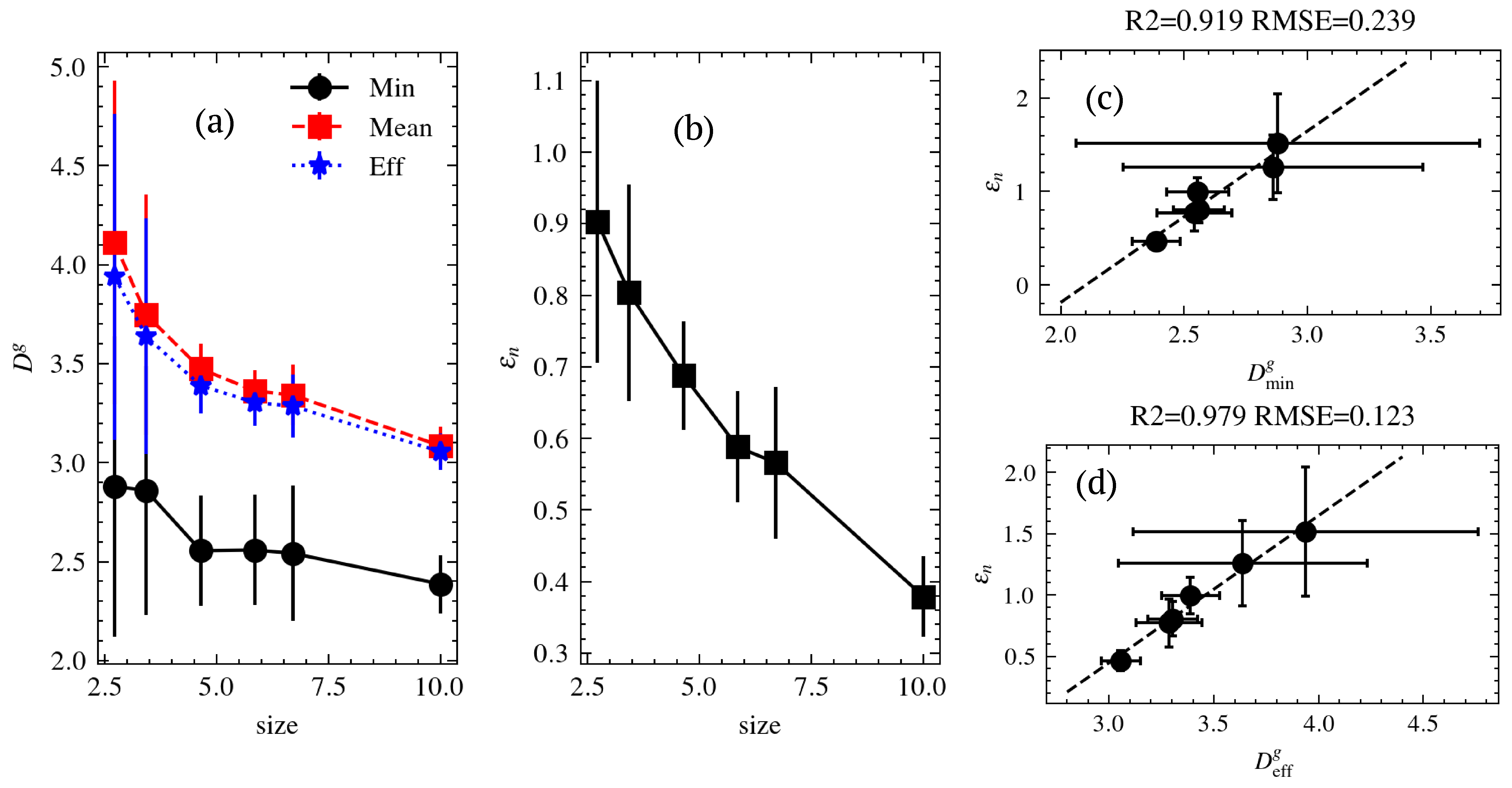}
  \caption{Lengthscale effect of fracture of the CG model in Fig. \ref{fig:deform_CG_sb} understood from the SPs. (a) Minimum, mean, and effective length of SPs when the length is increased.  (b) Strain at fracture nucleation continuously decreases as the system becomes larger. (c) Linear correlation between minimum SP length and fracture nucleation strain. The errorbars represent standard deviations over at least 10 independent runs (entire procedure). (d) The linear correlation between effective SP length and fracture nucleation strain. This correlation is stronger than that in (c), due to the necessity of considering all near-shortest SPs in large systems. }
  \label{fig:SP_SI_CG1}
\end{figure}

The second model we studied is similar to the one in Sec. \ref{sec:CGmodel_lh} but with a shorter bond length. It employs harmonic bonds with the equilibrium bond length $l_0=0.8$, the bond strength $K=100$, and the breakage length criterion $l_c=1.1$. As shown in Fig. \ref{fig:deform_CG_sb}, the lengthscale dependence of fracture is more pronounced than the one in Sec. \ref{sec:CGmodel_lh}. Notably, in the model with 50,000 beads, the fracture nucleates almost immediately after yielding and completes before a strain of 0.5, showing better agreement with the experimental observation. Similarly, we analyze the change of minimum, mean, and effective lengths of SPs as functions of system size, as shown in Fig. \ref{fig:SP_SI_CG1}. And the fracture nucleation strain shows a clear linear correlation with the effective SP length. 

\subsection{CG models with stiffer quartic bonds}

\begin{figure}[H]
  \centering
  \includegraphics[width=0.5\linewidth]{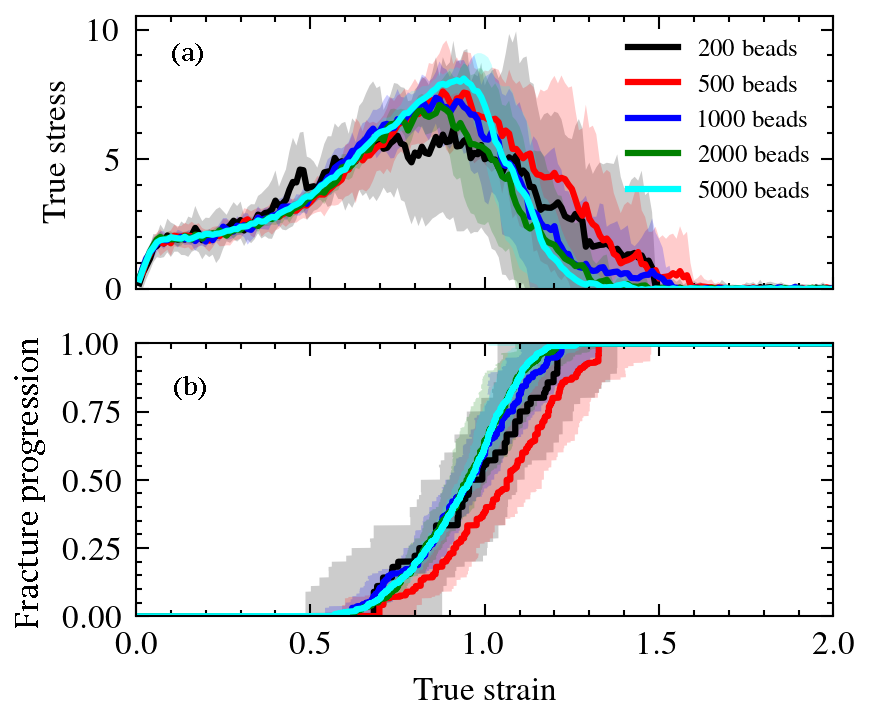}
  \caption{Same as in Fig. \ref{fig:deform_CG_sb} but for a different CG model with quartic bonds. The parameters for the bonds are given in Sec. \ref{sec:computational_models}. The shift of fracture due to increase of lengthscale is less evident in this model, possibly related to the increased bond stiffness for the quartic bonds. }
  \label{fig:deform_CG_quartic}
\end{figure}

The third model we studied employs quartic bonds, as detailed in Sec. \ref{sec:computational_models}. As shown in Fig. \ref{fig:deform_CG_sb}, the lengthscale dependence of fracture is less pronounced compared to the other two models presented above. We analyze the change of minimum, mean, and effective lengths of SPs as functions of system size, as shown in Fig. \ref{fig:SP_SI_CG1}. The minimum SP length converges very quickly but the effective length continuously decreases, showing a better linear correlation with the fracture nucleation strain. 

\begin{figure}[H]
  \centering
  \includegraphics[width=0.9\linewidth]{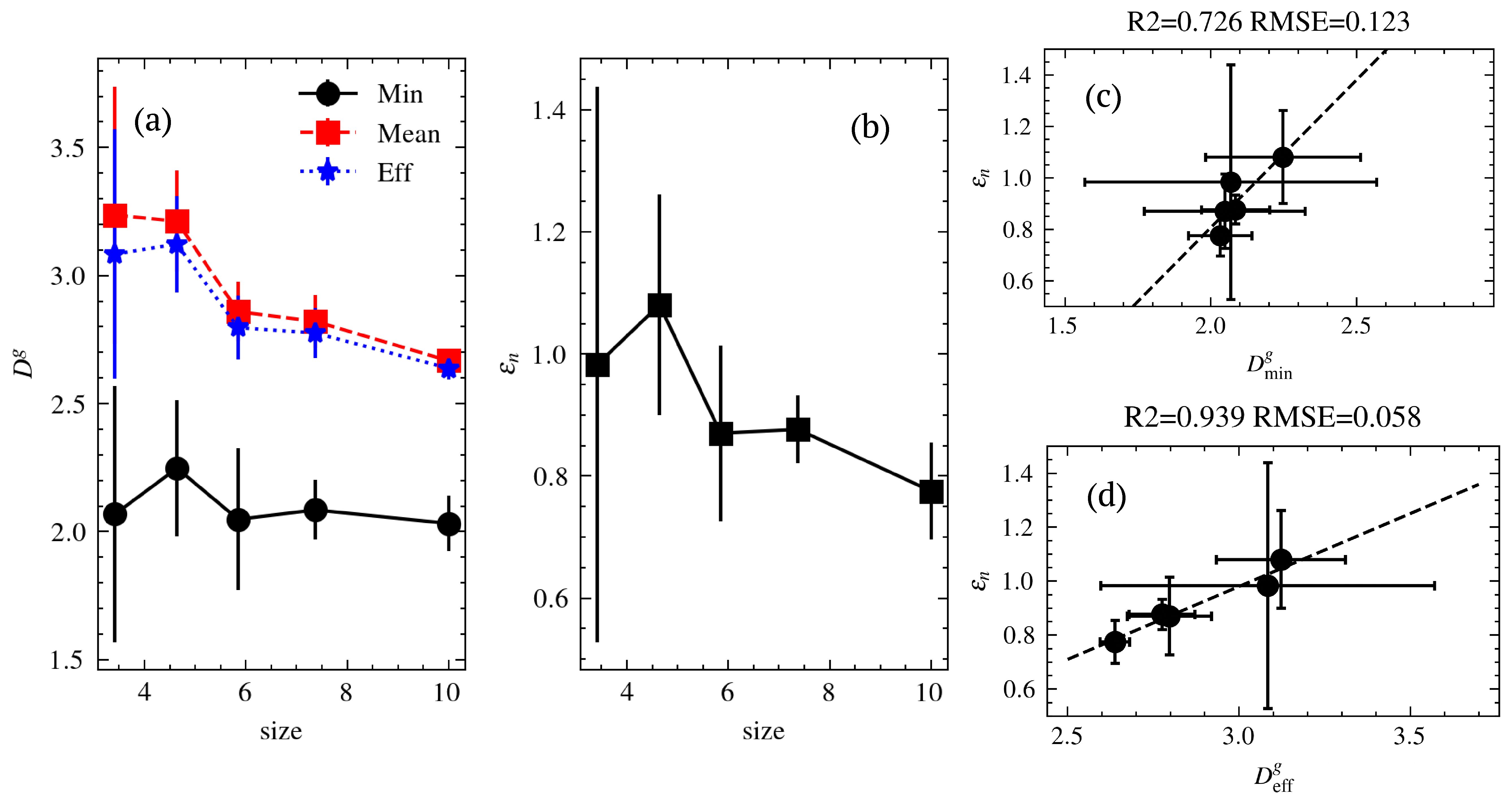}
  \caption{Lengthscale effect of fracture of the CG model in Fig. \ref{fig:deform_CG_quartic} understood from the SPs. The plots are similarly organized as those in Fig. \ref{fig:SP_SI_CG1}. Note that with increasing lengthscale, the minimum SP length does not change evidently in this case, but the fracture nucleation occurs earlier, which can be explained by the decreasing effective SP length. }
  \label{fig:SP_SI_quartic}
\end{figure}

\subsection{Network models}

\begin{figure}[H]
  \centering
  \includegraphics[width=0.8\linewidth]{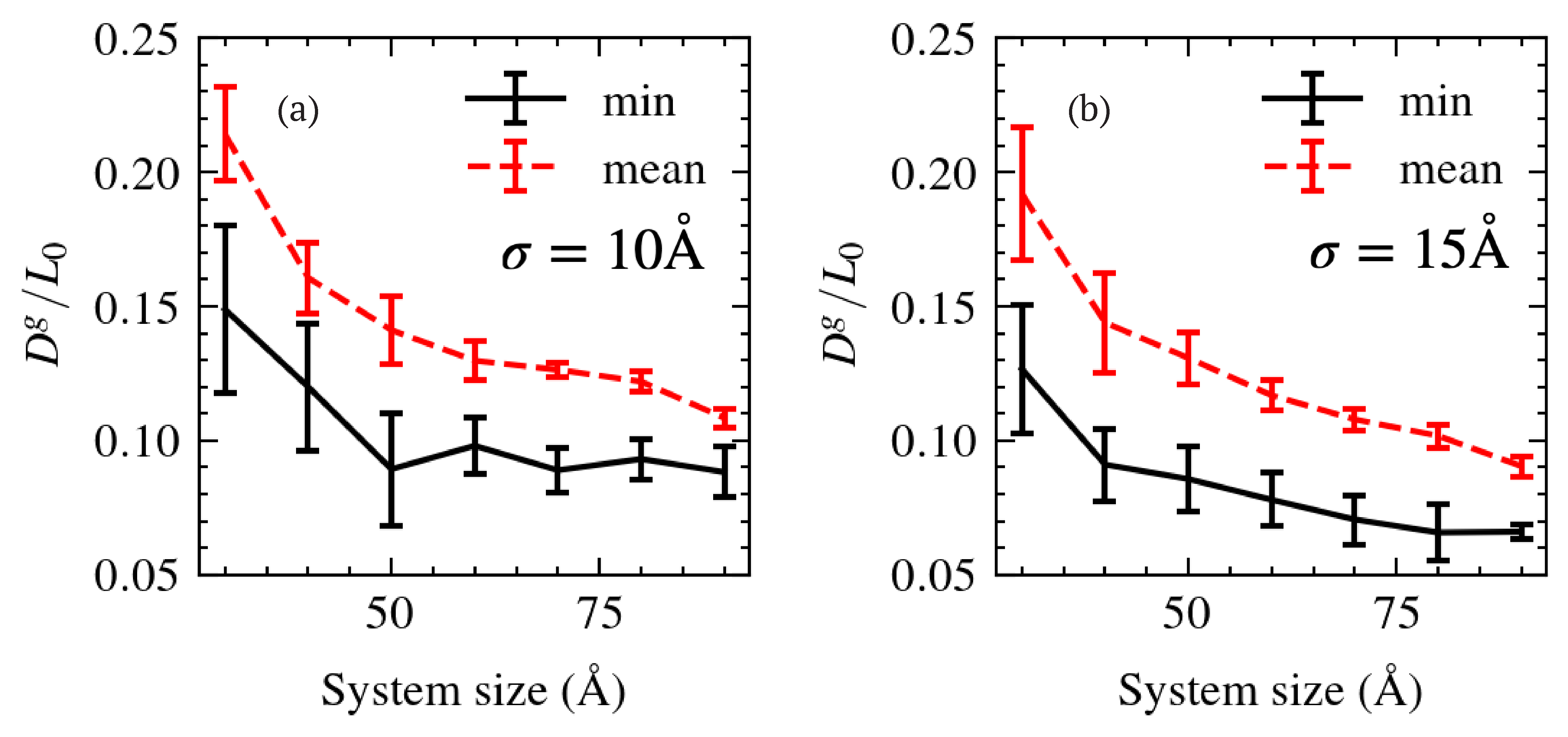}
  \caption{Minimum and mean of SP lengths at various system sizes in the network models for (a) $\sigma$ = 10 {\AA} and for (b) $\sigma$ = 15 {\AA}, where $\sigma$ is the standard deviation in Gaussian probability when linking the network, as introduced in Sec. \ref{sec:computational_models}. }
  \label{fig:network_model_size}
\end{figure}

To ensure the lengthscale dependence is broadly applicable to networks generated in different ways,  we also evaluate this in a random network model, which is connected based on a Gaussian probability over fixed node positions instead of reaction-like crosslinking in the all-atom and CG models, as detailed in Sec. \ref{sec:computational_models}. As shown in Fig. \ref{fig:network_model_size}A, the minimum SP length converges after some initial decline, whereas the mean SP length continues to decreases towards the minimum length, suggesting a narrower SP length distribution towards shorter lengths. This observation is consistent with those in the aforementioned CG models. When the standard deviation $\sigma$ in the Gaussian probability is increased, the size at convergence increases as well, about five times of the $\sigma$ value. This suggests that the polymer strand length, which is related to the $\sigma$ value, can also influence the fracture properties through SP characteristics. 

In general, these observations in different models are in qualitative agreement with each other, suggesting a potentially universal trend for lengthscale dependence of SP lengths in thermosetting networks regardless of simulation details.  

\subsection{Intuition for the effective lengths of shortest paths}\label{sec:effective_length}

In this subsection, we will explain why the SP effective length has the form of $\frac{\sum_i 1}{\sum_i (L_0/D_i)}$, similar to but slightly different from the the form of resistivity in parallel circuits. 

Suppose we have multiple paths that are all near-shortest,  they are all taut under strain close to fracture nucleation. If we let topological lengths be $D_i$ for the $i$th SP, the average length of each bond on the SP would be 
\begin{equation}
    d_i = L/D_i
\end{equation}
where $L=(\epsilon+1)\cdot L_0$ is the system length in the deformed direction and $L_0$ is the initial equilibrium length. 

We assume that the probability of bond breaking on a SP depending on the averaged bond length $d_i$ follows a Boltzmann distribution,
\begin{equation}
    P(d_i) \sim \exp(\frac{E - E_d}{k_B T}) \sim \exp(\frac{1/2K(d_i-d_0)^2-E_{d}}{k_B T})
\end{equation}
where $d_0$ is the equilibrium bond length, $K$ is the bond spring constant, $E_d$ is the bond dissociation energy, $T$ is the temperature, and $k_B$ is the Boltzmann constant. Since in reality bond breakage only occurs within a small range of length variation and if we assume it starts from $d'$, the probability above can be approximated using the Taylor expansion, 
\begin{equation}
    P(d_i) \sim P(d') + P(d') K (d_i-d_0) (d_i-d') \sim P(d') + P(d')K(d'-d_0)(d_i-d')
\end{equation}
As we can see, this probability is more like a quadratic function of $d_i$ but in a very narrow range it can be roughly treated as a linear function of $d_i$. For simplicity, let $P(d_i) \approx \alpha d_i +c$, where $\alpha$ and $c$ are fixed parameters depending on the system. 

In total, the probability of the system having at least one broken bond is 
\begin{equation}
    P_t = 1-\prod_i(1-P(d_i)) 
\end{equation}
where $i$ is still the index of SP. Note that $P_t$ at fracture nucleation is a small value because the attempted frequency (inverse to $P_t$) related to the frequency of vibrations is large. In this sense, 
\begin{equation}
    P_t \approx \sum_i P(d_i) = \sum_i [\alpha \cdot(L/D_i) +c]  =\alpha \sum_i \frac{(\varepsilon_n+1)L_0}{D_i} + cN_{SP}
\end{equation}
where $N_{SP} = \sum_i 1$ is the number of near-shortest SPs and $\varepsilon_n$ is the strain at fracture nucleation. Reorganizing this leads to 
\begin{equation}
    \varepsilon_n \approx \frac{P_t}{\alpha} \frac{1}{\sum_i (L_0/D_i)}-\frac{c}{\alpha} \frac{\sum_i 1}{\sum_i (L_0/D_i)} - 1
\end{equation}
Because $P_t$ is small and $\sum_i 1 \gg 1$, the second term will be dominant in fracture nucleation, and thus $D^g_\mathrm{eff} = \frac{\sum_i 1}{\sum_i (L_0/D_i)}$ works like the effective length of SPs. We should note that this effective treatment ideally only works for those near-shortest SPs in larger systems but it is not easy to identify exactly which of SPs belong to the ``near-shortest'', therefore, we compute $D^g_\mathrm{eff}$ based on the entire set of SPs and study the correlation between this approximated feature and fracture nucleation. The results are as discussed above, which satisfactorily explains the dependence on the lengthsacle.  

\section{Local stress distribution in the all-atom model}

\begin{figure}[H]
  \centering
  \includegraphics[width=0.8\linewidth]{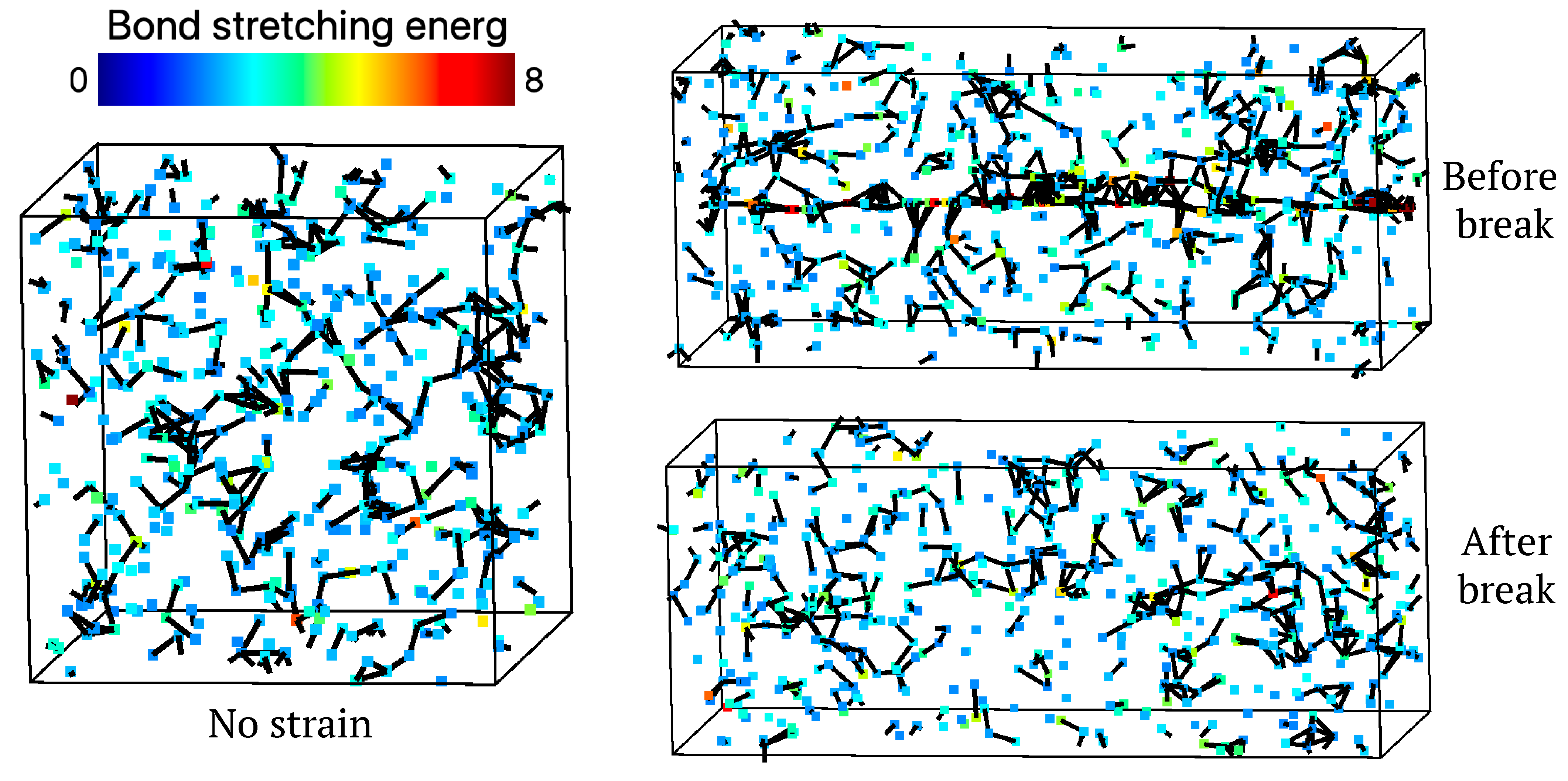}
  \caption{Local stress distribution in 3D corresponding to Fig. 1 in the main text. }
  \label{fig:local_stress_3d}
\end{figure}

\begin{figure}[H]
  \centering
  \includegraphics[width=0.8\linewidth]{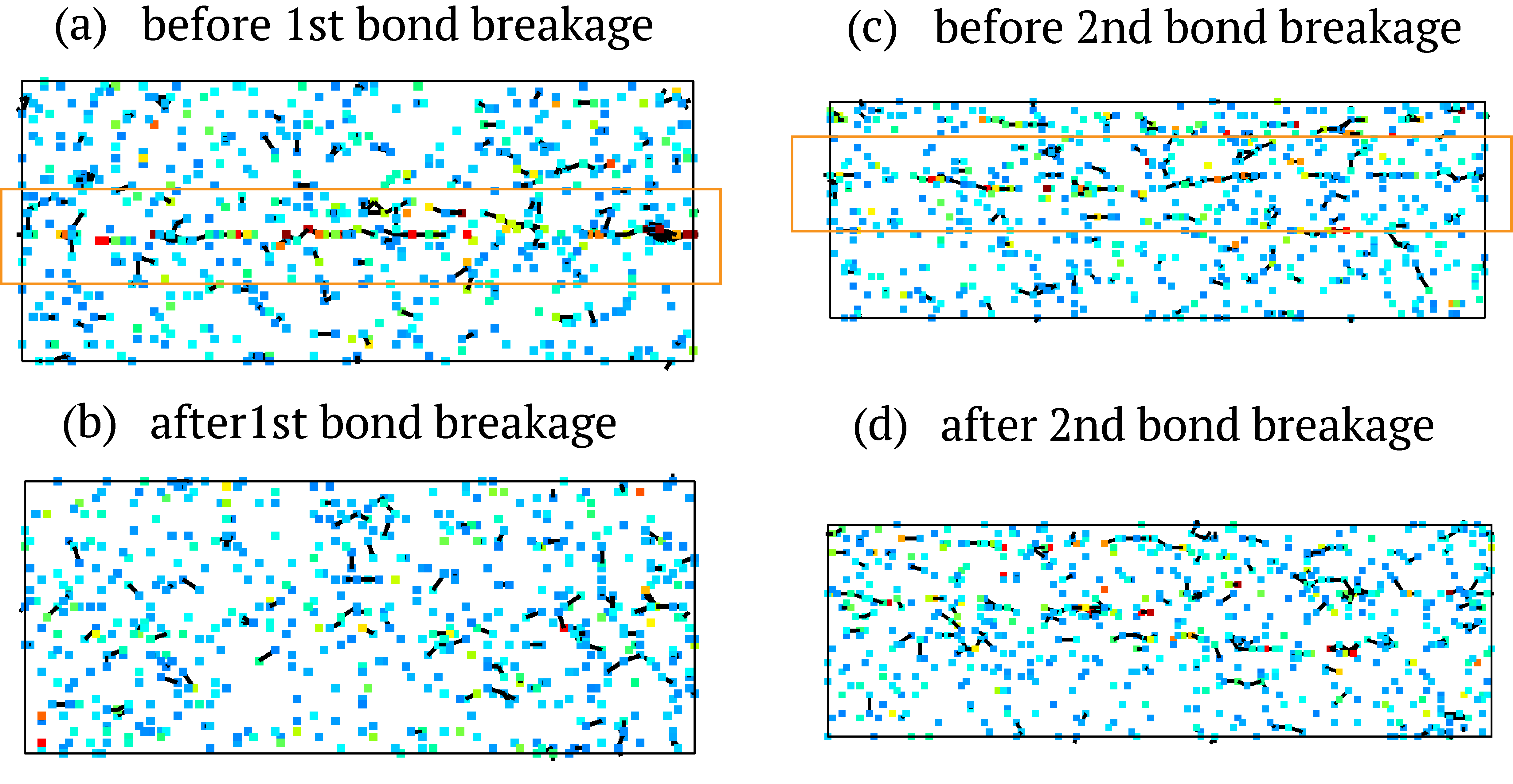}
  \caption{Local stress distribution around the first and second bond breakages in the MLABT simulation, respectively. Note that this is the same plot (for the first bond breakage) as in Fig. 1 but uses a smaller cutoff distance (4 {\AA} here) for connecting bonds by black line segments. The orange rectangles highlight the connected local stress pathways formed before bond breakage. }
  \label{fig:local_stress_smaller_cutoff}
\end{figure}

As described in the main text, regions under high stress align to form a linear pathway across the material in the direction of extension, immediately before the moment of fracture nucleation. This phenomenon can also be observed clearly from the 3D visualization presented in Fig. \ref{fig:local_stress_3d}. In addition, to ensure that the observed linear pathway is a genuine feature, instead of an artifact of the visualization settings, we present another plot (Fig. \ref{fig:local_stress_smaller_cutoff}) akin to Fig. 1 in the main text yet employing a reduced distance cutoff for indicating the local stressed region. The pathway is distinctly visible and this observation holds true for the second bond breakage as well. 

\bibliography{SP}